\documentclass[sn-mathphys-num, iicol]{sn-jnl}

\usepackage{graphicx}%
\usepackage{multirow}%
\usepackage{amsmath,amssymb,amsfonts}%
\usepackage{amsthm}%
\usepackage{mathrsfs}%
\usepackage[title]{appendix}%
\usepackage{xcolor}%
\usepackage{textcomp}%
\usepackage{manyfoot}%
\usepackage{booktabs}%
\usepackage{algorithm}%
\usepackage{algorithmicx}%
\usepackage{algpseudocode}%
\usepackage{listings}%
\usepackage{physics}
\usepackage[version=4]{mhchem}
\usepackage[super]{nth}
\usepackage{siunitx}
\usepackage{float}
\usepackage[font=scriptsize]{caption}
\usepackage{subcaption}
\usepackage{orcidlink}
\usepackage{placeins}

\theoremstyle{thmstyleone}%
\theoremstyle{thmstyletwo}%

\theoremstyle{thmstylethree}%

\begin{document}

\title[Article Title]{Solving coupled Non-linear Schrödinger Equations via Quantum Imaginary Time Evolution}


\author*{\fnm{Yang Hong} \sur{Li} \orcidlink{0009-0000-0431-8364}}\email{yhl51@cantab.ac.uk}

\author{\fnm{Jim} \sur{Al-Khalili} \orcidlink{0000-0003-3181-5280}}\email{j.al-khalili@surrey.ac.uk}
\equalcont{These authors contributed equally to this work.}

\author{\fnm{Paul} \sur{Stevenson} \orcidlink{0000-0003-2645-2569}}\email{p.stevenson@surrey.ac.uk}
\equalcont{These authors contributed equally to this work.}

\affil{\orgdiv{School of Mathematics and Physics, Faculty of Engineering and Physical Sciences}, \orgname{University of Surrey}, \orgaddress{\city{Guildford}, \postcode{GU2 7XH}, \country{UK}}}




\abstract{Coupled non-linear Schr\"{o}dinger equations are crucial in describing dynamics of many-particle systems. We present a quantum imaginary time evolution (ITE) algorithm as a solution to such equations in the case of nuclear Hartree-Fock approach. Under a simplified Skyrme interaction model, we calculate the ground state energy of an oxygen-16 nucleus and demonstrate that the result is in agreement with the classical ITE algorithm. We examine bottlenecks and deficiencies in the quantum algorithm and suggest possible improvements.}

\maketitle
\definecolor{DarkGreen}{RGB}{0,150,0}

\section{Introduction}

The dynamics of non-relativistic quantum systems are described by the Schr\"{o}dinger equation. For many-particle systems the effective potential is often dependent on the particle configuration itself. For example, this is the case in the Hartree-Fock approach, which is the basic method for reduction of the many-body problem to an approximate set of one-body problems, and from which higher-order approximations are built \cite[equation (3.15) and surrounding discussion]{JSuhonen.2007} In such a case, the Schr\"{o}dinger equation governing the system becomes a set of coupled non-linear differential equations. Finding the solution to such equations, and finding improved algorithms for doing so is both a long-standing and contemporary challenge in classical computing \cite{JChemPhys.55.2408,JComputPhys.27.221,PhysRevE.110.055304}, with new algorithms following new technologies \cite{SC20.1,IntJHighPerformComputAppl.30.85,JChemPhys.159.104101,arXiv:2401.00019}. Given the role of the Hartree-Fock basis for further computation, solution of such equations on a quantum computer fall under title of the state preparation problem \cite{IntJQuantumChem.122.e26853}. Here we explore the possibility of solution of coupled non-linear Schr\"odinger equations on a quantum computer, where advantages such as the exponential scaling of Hilbert space with qubit number, and quantum entanglement of qubits, could prove a useful aid to solution of the equations. This follows up on our previous study \cite{PhysRevC.109.044322} in which a simplified case involving only a single uncoupled Schr\"odinger equation was studied. The novelty in the present paper is to consider a more general case where the underlying physics problem gives rise to coupled non-linear Schr\"odinger equations and to study the extension needed to our quantum algorithm. While we specialise our solution to a nuclear physics case, we note that the possibility of using quantum computers to solve general differential equations is a wider active field of application of this emerging technology \cite{PhysRevLett.103.150502,SciRep.14.20156}.

One major advantage of quantum computers over classical computers is their ability to simulate other quantum systems \cite{IntJTheorPhys.21.467}. For instance, wave functions of many-particle systems can be more efficiently encoded by qubits than by bits. This idea has been widely used in fields like quantum chemistry 
\cite{RevModPhys.92.015003,ChemRev.119.10856,Nature.618.500,Science.369.1084}, materials \cite{NatComputSci.2.424,QuantumSciTechnol.6.043002}, and nuclear physics \cite{TESNAT2015.01008,ChinPhysB.30.020306,IntJUnconvComput.18.83,AdvQuantumTechnol.2300219}.

We previously presented an approach to solving the nuclear Hartree-Fock equation for \ce{^{4}He} via quantum imaginary time evolution (ITE) \cite{PhysRevC.109.044322}, which involves finding the solution of a non-linear Schr\"{o}dinger equation. In this paper, we give an expansion of the previous algorithm for coupled non-linear Schr\"{o}dinger equations. Using the \ce{^{16}O} nucleus as an example, we show that our implementation gives identical results as those from the classical ITE algorithm, while also demonstrating a method for solving a set of coupled non-linear differential equations.

\section{Imaginary Time Evolution}

The time evolution of a state $\ket{\psi\left(\vec{r},t\right)}$ under the Hamiltonian $\hat{H}$ is described by the time-dependent Schr\"{o}dinger equation (TDSE) 
\begin{equation}
    \hat{H}\ket{\psi\left(\vec{r},t\right)}=i\hbar\frac{\partial}{\partial t}\ket{\psi\left(\vec{r},t\right)}\text{,}
\end{equation}
which can be re-written in imaginary time ($t\rightarrow-i\tau$) as
\begin{equation}
    \frac{\partial}{\partial\tau}\ket{\psi\left(\vec{r},\tau\right)}=-\frac{\hat{H}}{\hbar}\ket{\psi\left(\vec{r},\tau\right)}\text{.}
\end{equation}
Given an initial state $\ket{\psi\left(\vec{r},0\right)}$, the time evolution gives the general solution
\begin{equation}\label{im_time}
    \ket{\psi\left(\vec{r},\tau\right)}=\mathcal{N}\mathcal{T}\exp(-\int_0^\tau\frac{\hat{H}}{\hbar}d\tau')\ket{\psi\left(\vec{r},0\right)}\text{,}
\end{equation}
where $\mathcal{T}$ is the time-ordering operator and $\mathcal{N}$ is an operator that renormalizes the state after the application of the non-unitary imaginary time evolution operator. When $\tau\rightarrow\infty$, $\ket{\psi\left(\vec{r},\tau\right)}$ converges to the ground state of $\hat{H}$, provided that the initial state $\ket{\psi\left(\vec{r},0\right)}$ is not orthogonal to it \cite{PhysRevC.109.044322, JComputPhys.221.148}.

For a time-dependent Hamiltonian $\hat{H}(\tau')$, the integral in equation (\ref{im_time}) can be approximated by
\begin{equation}\label{IntApprox}
    -\int_0^\tau\frac{\hat{H}(\tau')}{\hbar}d\tau'\approx-\frac{\Delta\tau}{\hbar}\sum_{k=0}^{k_{total}}\hat{H}(k\Delta\tau)\text{,}
\end{equation}
where $\Delta\tau=\frac{\tau}{k_{total}}$ is the imaginary time step and $k_{total}$ is the total number of steps. Although the Hamiltonian we use (see section \ref{O16}) is independent of real time, it is nonlinear and we deal with the nonlinearity through an iterative approach in which the Hamiltonian converges from an initial guess to the true Hamiltonian as the solution is reached. Hence we have an iteration-dependent Hamiltonian, which manifests in the imaginary time approach as an imaginary time-dependent Hamiltonian. (see Appendix \ref{NlH}) Using the first order Suzuki-Trotter decomposition \cite{CommunMathPhys.51.183} with implied time-ordering
\begin{equation}
    e^{\delta(A+B)}=e^{\delta A}e^{\delta B}+O(\delta^2)\text{,}
\end{equation}
equation (\ref{im_time}) can be rewritten as
\begin{equation}
    \ket{\psi\left(\vec{r},\tau\right)}=\prod_{k=0}^{k_{total}}\hat{O}^{(k)}\left(\Delta\tau\right)\ket{\psi\left(\vec{r},0\right)}\text{,}
\end{equation}
where the non-unitary ITE operator\footnote{In our previous work \cite{PhysRevC.109.044322} we used the notation $\hat{U}$ for this operator. We realised that this might cause confusion about the operator's unitarity and have now updated the notation to $\hat{O}$.} $\hat{O}^{(k)}$ is given by
\begin{equation}
    \hat{O}^{(k)}\left(\Delta\tau\right)=\mathcal{N}\exp[-\frac{\Delta\tau}{\hbar}\hat{H}(k\Delta\tau)]\text{.}
\end{equation}

With a sufficiently small imaginary time step $\Delta\tau\ll1$, $\hat{O}^{(k)}$ can be approximated by
\begin{equation}\label{eq:uapprox}
    \hat{O}^{(k)}(\Delta\tau)\approx\mathcal{N}\left[1-\frac{\Delta\tau}{\hbar}\hat{H}(k\Delta\tau)\right]\text{.}
\end{equation}

\section{\ce{^{16}O} Nuclear Model}\label{O16}
In our previous work we calculated the ground state of \ce{^{4}He} with a simplified Skyrme interaction \cite{PhysRevC.109.044322, PhysRevC.60.044302, NuclPhys.9.615, ComputPhysCommun.102.166} as our test model. In this work we will assume the same simplified effective interaction with the radial Hartree-Fock Hamiltonian
\begin{equation}\label{Ham}\begin{aligned}
    \hat{H}_{\mathrm{hf}}^{l}=T^l+V^l&=-\frac{\hbar^2}{2m}\left[\frac{d^2}{dr^2}-\frac{l\left(l+1\right)}{r^2}\right]\\
    &\quad+\left[\frac{3}{4}t_0\rho\left(r\right)+\frac{3}{16}t_3\rho^2\left(r\right)\right]\text{,}
\end{aligned}\end{equation}
where $t_0$ and $t_3$ are Skyrme force parameters. The density $\rho\left(r\right)$ is given by
\begin{equation}\label{density}
    \rho\left(r\right)=\frac{1}{4\pi r^2}\sum_{n,l}4\left(2l+1\right)\left|u_{nl}\right|^2 \text{,}
\end{equation}
where $u_{nl}\left(r\right)/{r}$ are the radial component of the single particle (SP) wave functions $\varphi_{nlmm_sq}\left(\vec{r},s,I_3\right)$. Spin and isospin degeneracies are accounted for with a factor of 4.

In \ce{^{4}He} all four nucleons are in the $0s_{1/2}$ state (in usual nuclear spectroscopic notation \cite[chapter 3]{JSuhonen.2007}, where 0 is the principal quantum number, $s$ corresponds to orbital angular momentum $l=0$, and $1/2$ is the total angular momentum $j$), and there is only one non-linear Schr\"{o}dinger-like equation to solve. In the \ce{^{16}O} case, the $0s_{1/2}$ ($l=0$), $0p_{1/2}$ ($l=1$), and $0p_{3/2}$ ($l=1$) states are all fully occupied. Since the $0p_{1/2}$ and $0p_{3/2}$ states share the same radial spatial wave function, this corresponds to two coupled non-linear Schr\"{o}dinger like equations. The Hamiltonians for the $s$ states and $p$ states differ in the kinetic term,
\begin{equation}\begin{aligned}
    T^0&=-\frac{\hbar^2}{2m}\frac{d^2}{dr^2}\\
    T^1&=-\frac{\hbar^2}{2m}\left(\frac{d^2}{dr^2}-\frac{2}{r^2}\right)\text{,}
\end{aligned}\end{equation}
but share the same potential term
\begin{equation}
    V^0=V^1=V(r)=\frac{3}{4}t_0\rho\left(r\right)+\frac{3}{16}t_3\rho^2\left(r\right)\text{,}
\end{equation}
where
\begin{equation}
    \rho\left(r\right)=\frac{1}{\pi r^2}\left(\left|u_{00}\right|^2+3\left|u_{01}\right|^2\right)\text{.}
\end{equation}
Hence we can write the radial Hatree-Fock equations for both $s$ and $p$ states,
\begin{align}
    \left[-\frac{\hbar^2}{2m}\frac{d^2}{dr^2}+V(r)\right]u_{00}&=\varepsilon_{00}u_{00}\text{,}\\
    \left[-\frac{\hbar^2}{2m}\frac{d^2}{dr^2}+\frac{\hbar^2}{mr^2}+V(r)\right]u_{01}&=\varepsilon_{01}u_{01}\text{,}
\end{align}
where $\varepsilon_{0l}$ are SP energies for the corresponding states.

These two equations are coupled explicitly through the appearance of both $u_{00}$ and $u_{01}$ in the potential $V(r)$.

\section{Quantum Implementation}
We encode our states in the 3D isotropic oscillator basis (with oscillator length $\frac{1}{b}$ and corresponding angular momentum $l$) \cite{PhysRev.177.1519,ComputPhysCommun.200.220,SciChinaPhysMechAstron.66.240311,ComputPhysCommun.102.166}
\begin{equation}\label{exp}
    u_{nl}\left(r\right)=\sum_{n'}\alpha_{nln'}\mathcal{R}^{b, l+\frac{3}{2}}_{n'}\left(r\right),\;\; \alpha_{nln'}\in\mathbb{R}\:\forall\:n,l,n'\text{.}
\end{equation} 
The $2^N$ expansion coefficients $\alpha_{nln'}$ are stored in a $N$-qubit target state $\ket{\psi_{l}}$, where \begin{equation}\label{decomstate}
    \ket{{\psi_{l}}}=\sum_{n'=0}^{2^{N}-1}\alpha_{0ln'}\ket{n'}\text{,}
\end{equation}
and $\ket{0}=\ket{00\ldots00}$, $\ket{1}=\ket{00\ldots01}$, $\ket{2}=\ket{00\ldots10}$ \textit{etc}. Matrix elements of the density in the basis are then computed as a sum of integrals of the product of four oscillator radial wave functions $\mathcal{R}^{b, l+\frac{3}{2}}_{n'}\left(r\right)$, \cite{PhysRevC.109.044322}
\begin{equation}
    \rho^l\sim\sum_j\int\frac{dr}{r^2}\mathcal{R}^l\mathcal{R}^l\mathcal{R}^j\mathcal{R}^j\text{,}
\end{equation}
where the superscript $l$ of $\rho^l$ indicates the sub-basis that $\rho(r)$ is expanded in. These calculated integrals are tabulated at the beginning and used in each step of the ITE.

Once the matrix form of the pre-normalised ITE operator $\hat{O}^{(k)}$ is obtained, it is decomposed into a sum of products of Pauli matrices (and identity matrices) acting on individual qubits \cite{arxiv:2111.00627}. This non-unitary Hermitian gate is implemented using the idea of a duality computer \cite{CommunTheorPhys.45.825,Res.2020.1}, with the aid of $2N$ auxiliary qubits and following the approach outlined in our previous work \cite{PhysRevC.109.044322}.

\begin{figure}[tbh]
    \centering
    \includegraphics[width=\columnwidth]{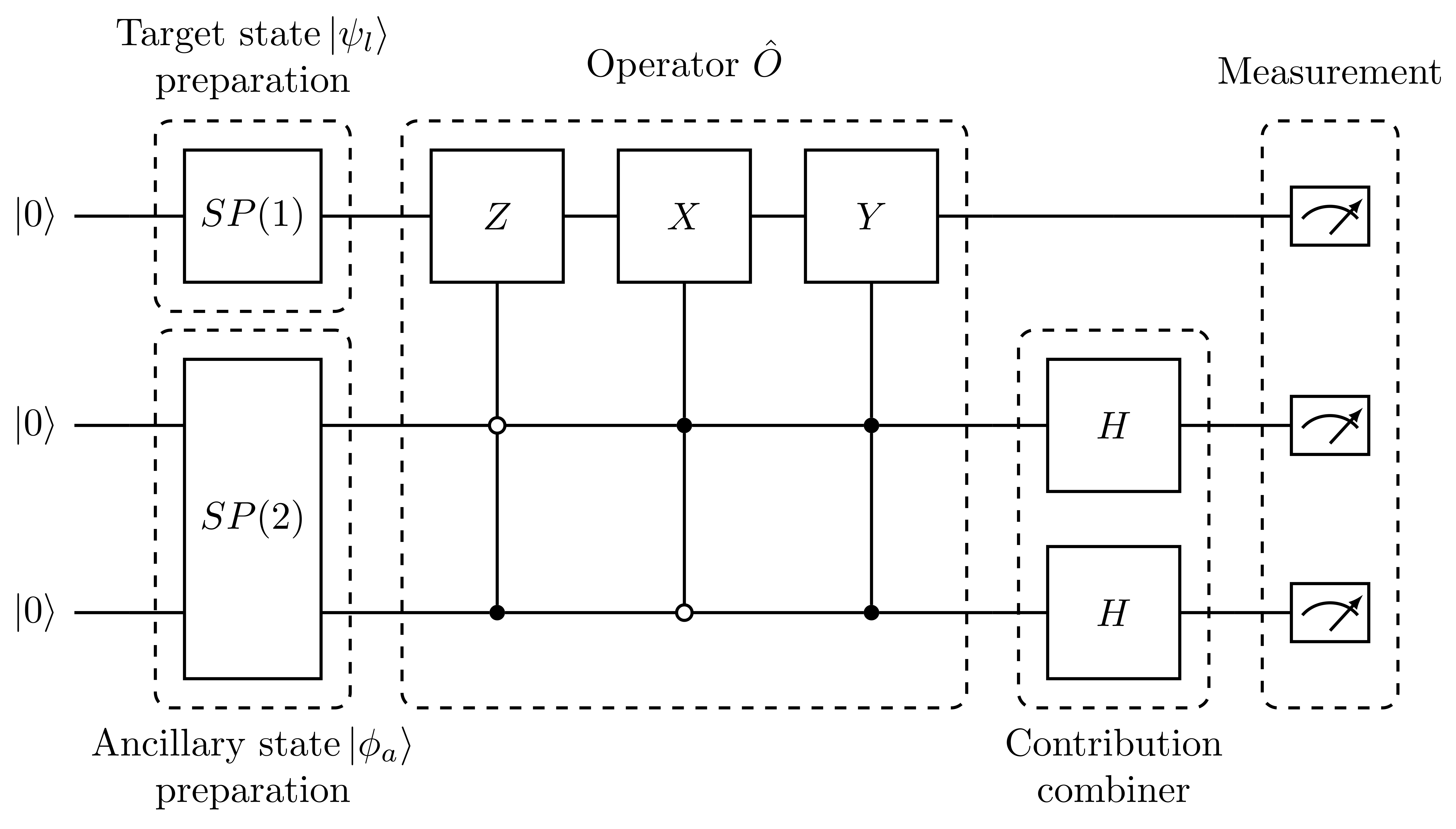}
    \caption{Quantum circuit for an ITE step in $N=1$ case. All ancillary qubits are required to be measured to be zero for a succesful evolution.} The state preparation subcircuits $SP(\mathfrak{N})$ are given in Appendix \ref{SP}.
    \label{QCN1}
\end{figure}

\section{Results}
We applied the approach above, using a classical algorithm and on a quantum simulator separately, to calculate the ground state energy of \ce{^{16}O} \cite{JChemPhys.148.094101}, for $N=1$ (2 expansion states) up to $N=4$ (16 expansion states) cases. We use the values $t_0=-1132.4\,\si{\MeV\femto\m\cubed}$ and $t_3=23610.4\,\si{\MeV\femto\m\tothe{6}}$ for the Skryme force parameters, and choose an imaginary time step of $\frac{\Delta\tau}{\hbar}=0.005\,\si{\MeV\tothe{-1}}$ \cite{PhysRevC.109.044322}. The values for the Skyrme force parameters were taken from a previous study \cite{PhysRevC.60.044302} in which they were adjusted to give the correct binding energy and radius of $^{16}$O at the Hartree-Fock level. The oscillator length $\frac{1}{b}$ for the basis is optimized by minimizing the calculated ground state energy. For each case, we start from two initial trial states of equal amplitudes for the oscillator states,
\begin{equation}
    u_{00}^{(0)}=\frac{1}{\sqrt{2^N}}\sum_{n'=0}^{2^N-1}\mathcal{R}^{{b}, \frac{3}{2}}_{n'}
\end{equation}
and
\begin{equation}
    u_{01}^{(0)}=\frac{1}{\sqrt{2^N}}\sum_{n'=0}^{2^N-1}\mathcal{R}^{{b}, \frac{5}{2}}_{n'}\text{.}
\end{equation}
The ITE is then performed for 400 iterations.

\begin{figure}[tbh]
    \centering
    \includegraphics[width=\columnwidth]{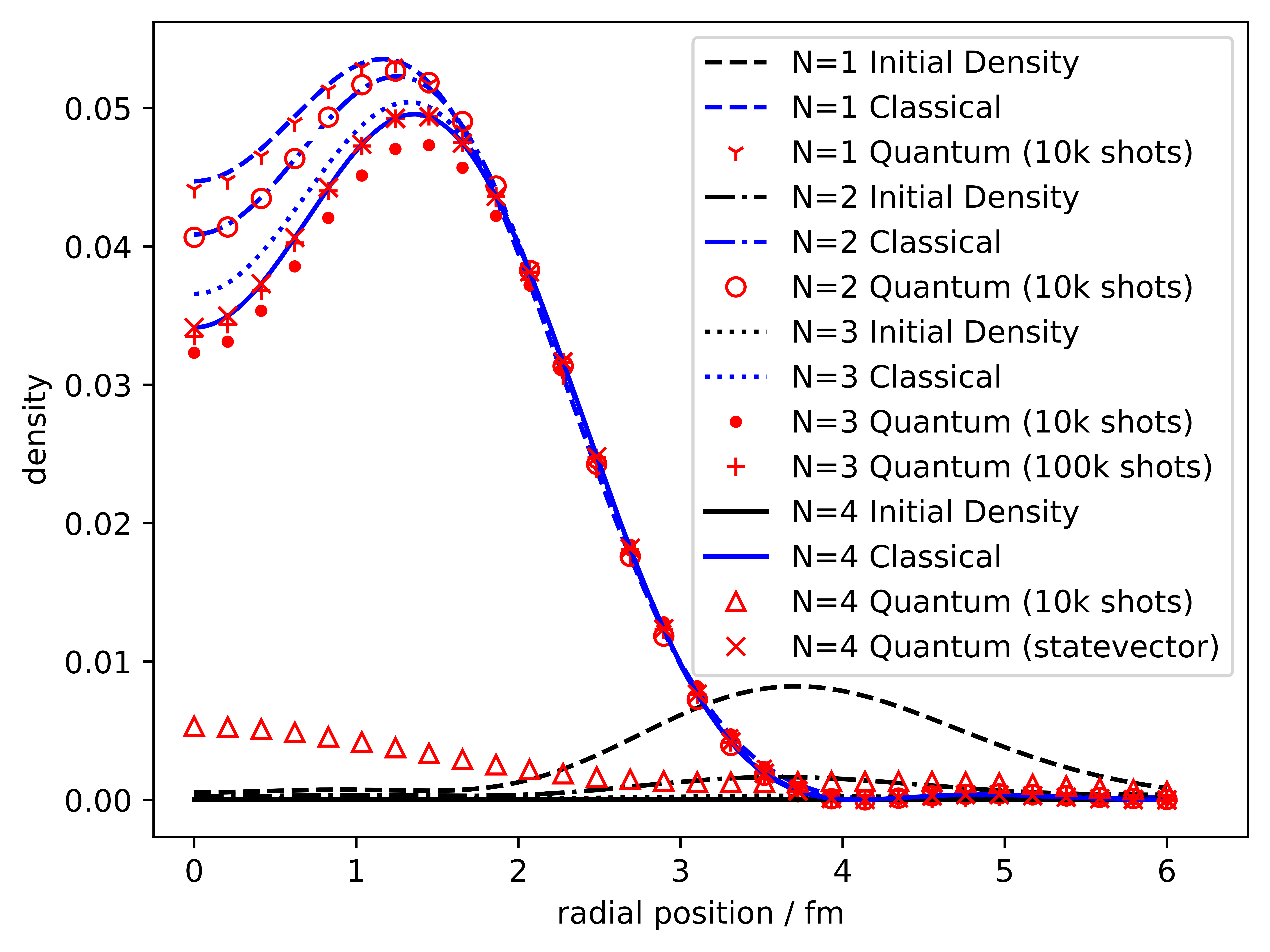}
    \caption{ITE after 400 iterations.}
    \label{OCQ1234_400}
\end{figure}

The quantum ITE procedure is implemented on a classically simulated quantum computer using the QASM backend from QISKIT \cite{PhysRevC.109.044322, MTreinish.2023} using 10000 shots per measurement. Figure \ref{OCQ1234_400} shows the final densities obtained by the ITE runs, while table \ref{gse_tab} and figure \ref{OCQ1234_En} show the ground state energy obtained.

\setlength{\tabcolsep}{3pt}
\begin{table}[tbh]
    \centering
    \begin{tabular}{cccccc} 
        \hline\hline
        \multirow{2}{*}{$N$}&\multirow{2}{*}{$\frac{1}{b}/\si{\femto\m}$} &\multicolumn{4}{c}{$E_{gs}/\si{\MeV}$}\\
        &&classical&10k shots&100k shots&statevector\\ \hline
        1&2.0402&-86.31&-86.29&-86.29&-86.31\\ 
        2&2.7040&-104.32&-103.85&-104.44&-104.32\\ 
        3&3.7060&-110.57&-106.19&-109.47&-110.57\\ 
        4&5.1508&-112.65&\textcolor{red}{11.91}&&-112.65\\ \hline\hline
    \end{tabular}
    \caption{Optimised oscillator length and \ce{^{16}O} Hartree-Fock ground state energy after 400 ITE iterations. Value obtained from failed convergence highlighted in red. The 100000 shots result for $N=4$ is not available due to impractical runtime.}
    \label{gse_tab}
\end{table}

\begin{figure}[tbh]
    \subfloat[\label{OCQ1234_En_full}]{\includegraphics[width=\columnwidth]{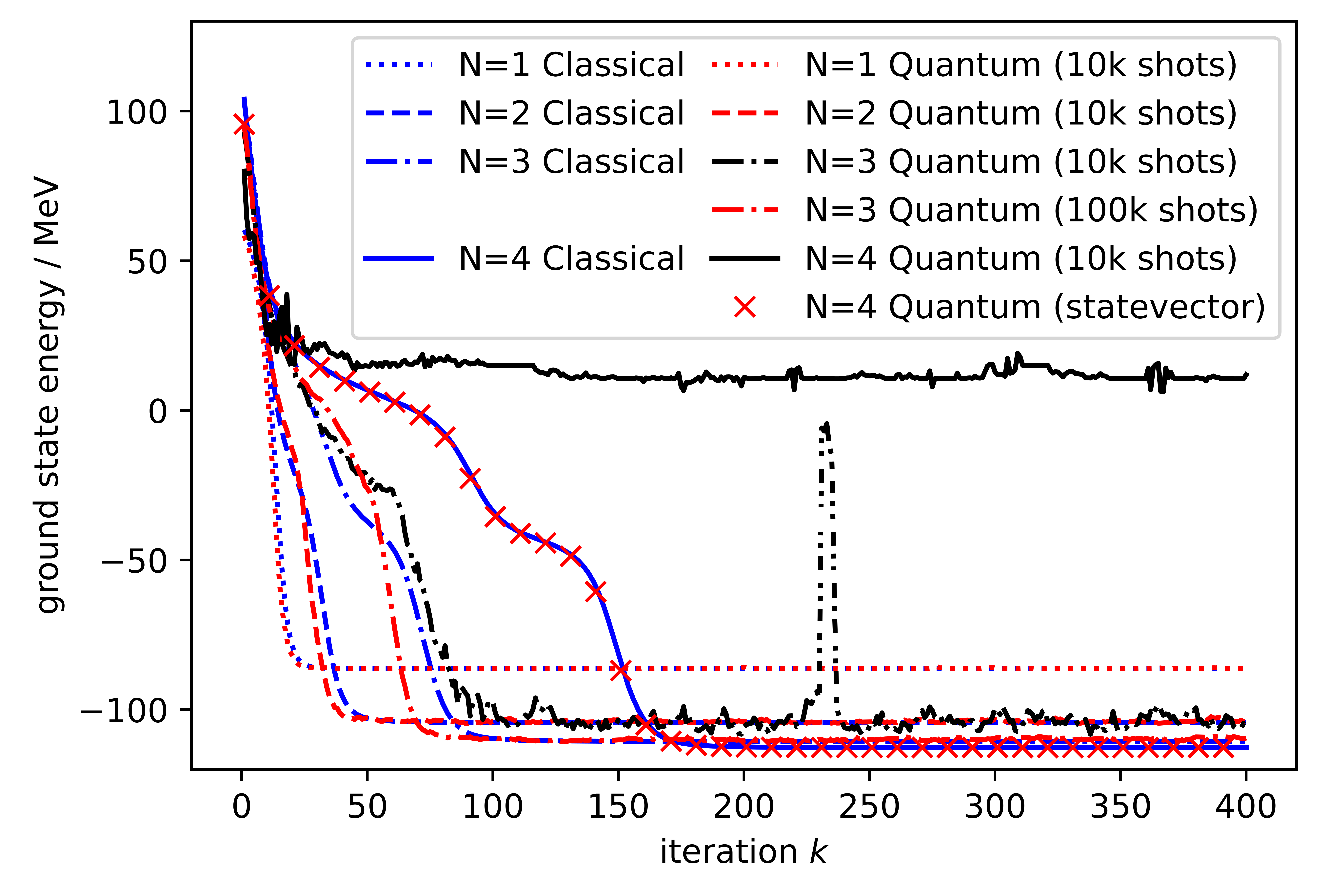}}
\end{figure}

\begin{figure}[tbh]
    \ContinuedFloat    
    \subfloat[\label{OCQ1234_En_zoom}]{\includegraphics[width=\columnwidth]{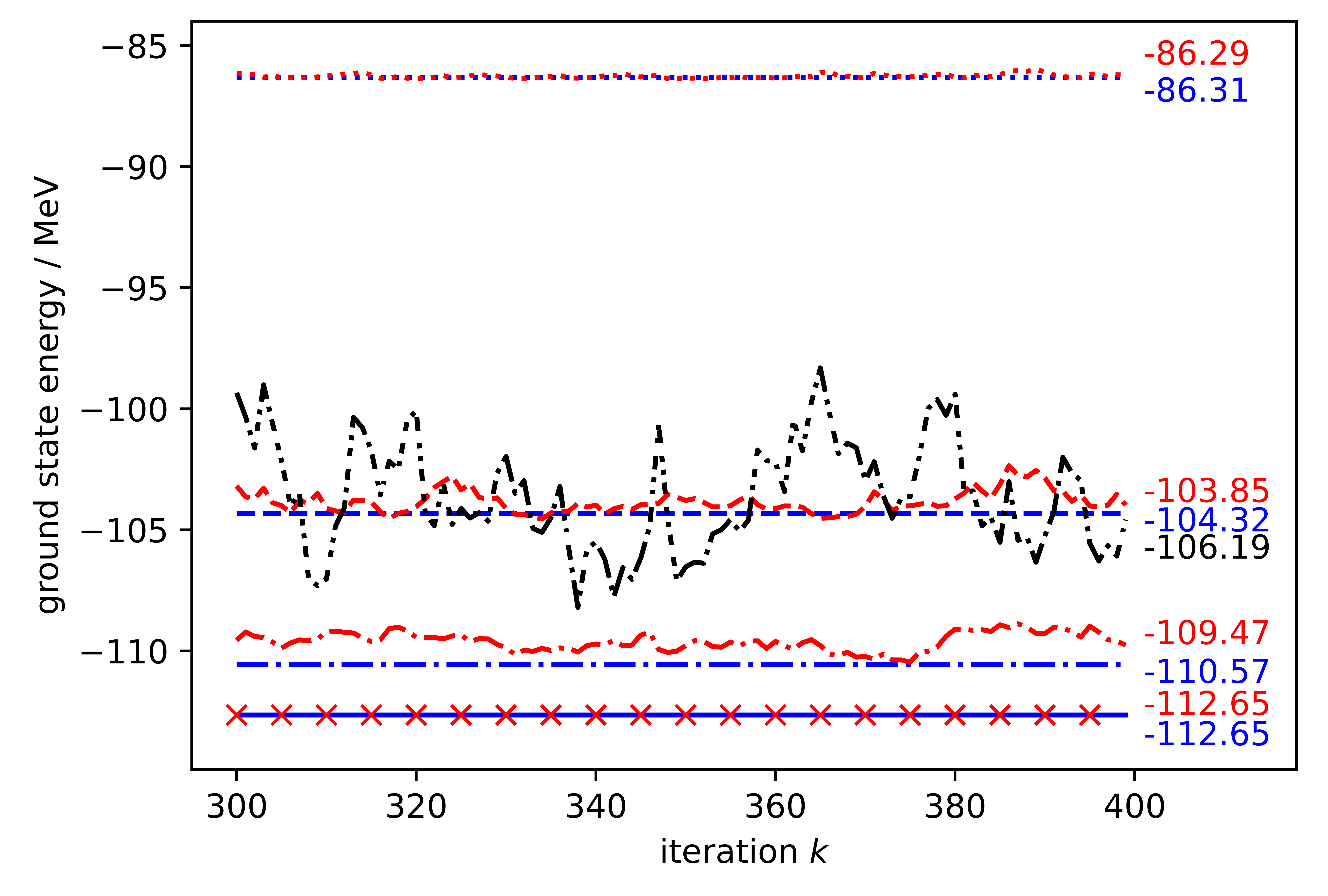}}
    \caption{Per iteration development of the ground state energy $E_{gs}$. (a) Full view. (b) Zoomed in view.}
    \label{OCQ1234_En}
\end{figure}

In the $N=1,2$ cases, the simulated quantum algorithm results are in close agreement with the classical implementation. Starting from $N=3$, the fluctuation $E_{gs}$ exhibits is greater than in the $N=1,2$ cases when using 10000 shots per measurement.

Figure \ref{O3} shows the evolution of the density as a function of iteration under the classical algorithm and quantum algorithm (10000 shots). This provides a closer look at the difference between the expected and actual behaviour of the algorithm.

\begin{figure}[tbh]
    \subfloat[\label{OC3}]{\includegraphics[width=\columnwidth]{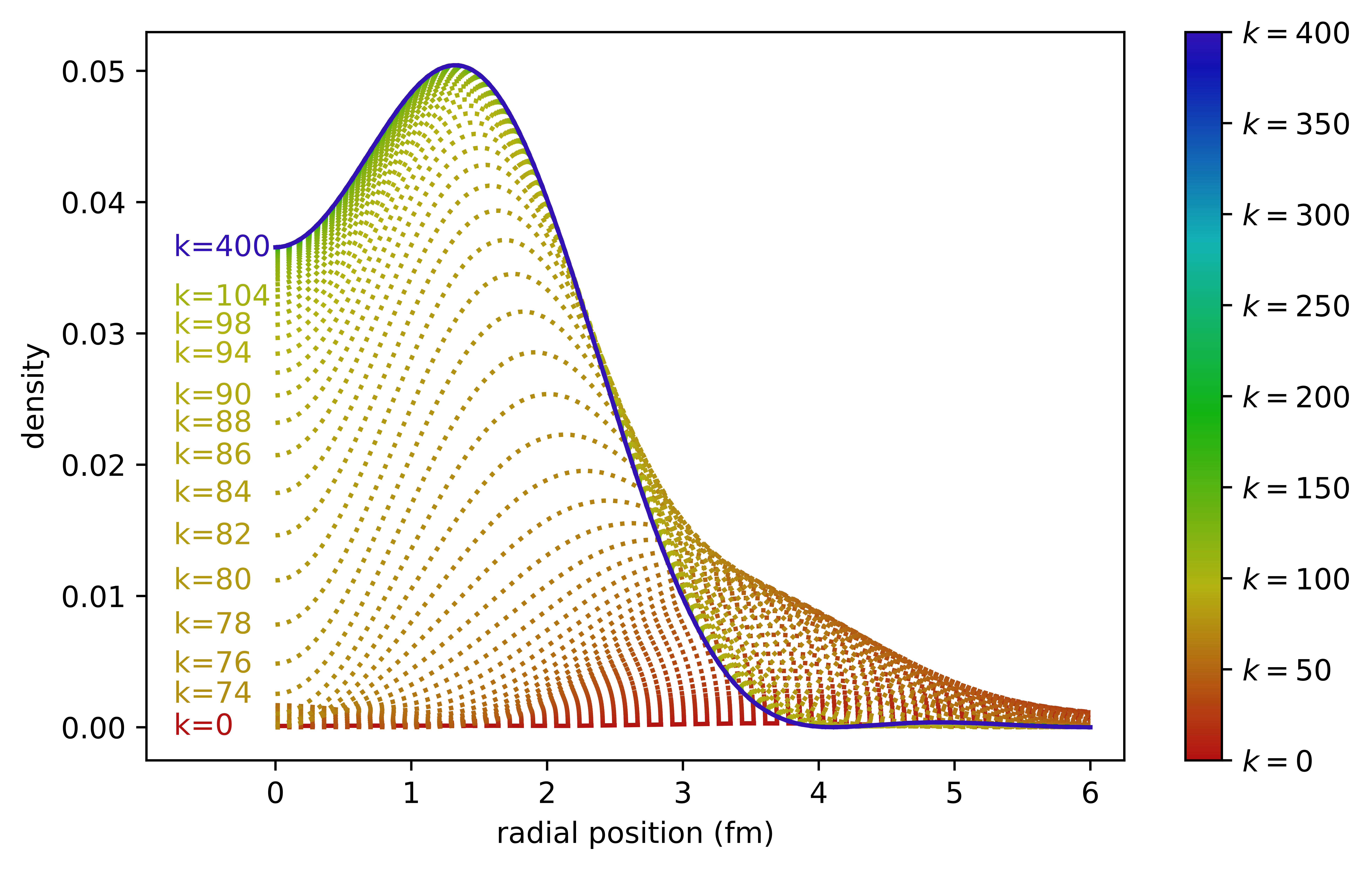}}
\end{figure}

\begin{figure}[H]
    \ContinuedFloat
    \subfloat[\label{OQm3_10k}]{\includegraphics[width=\columnwidth]{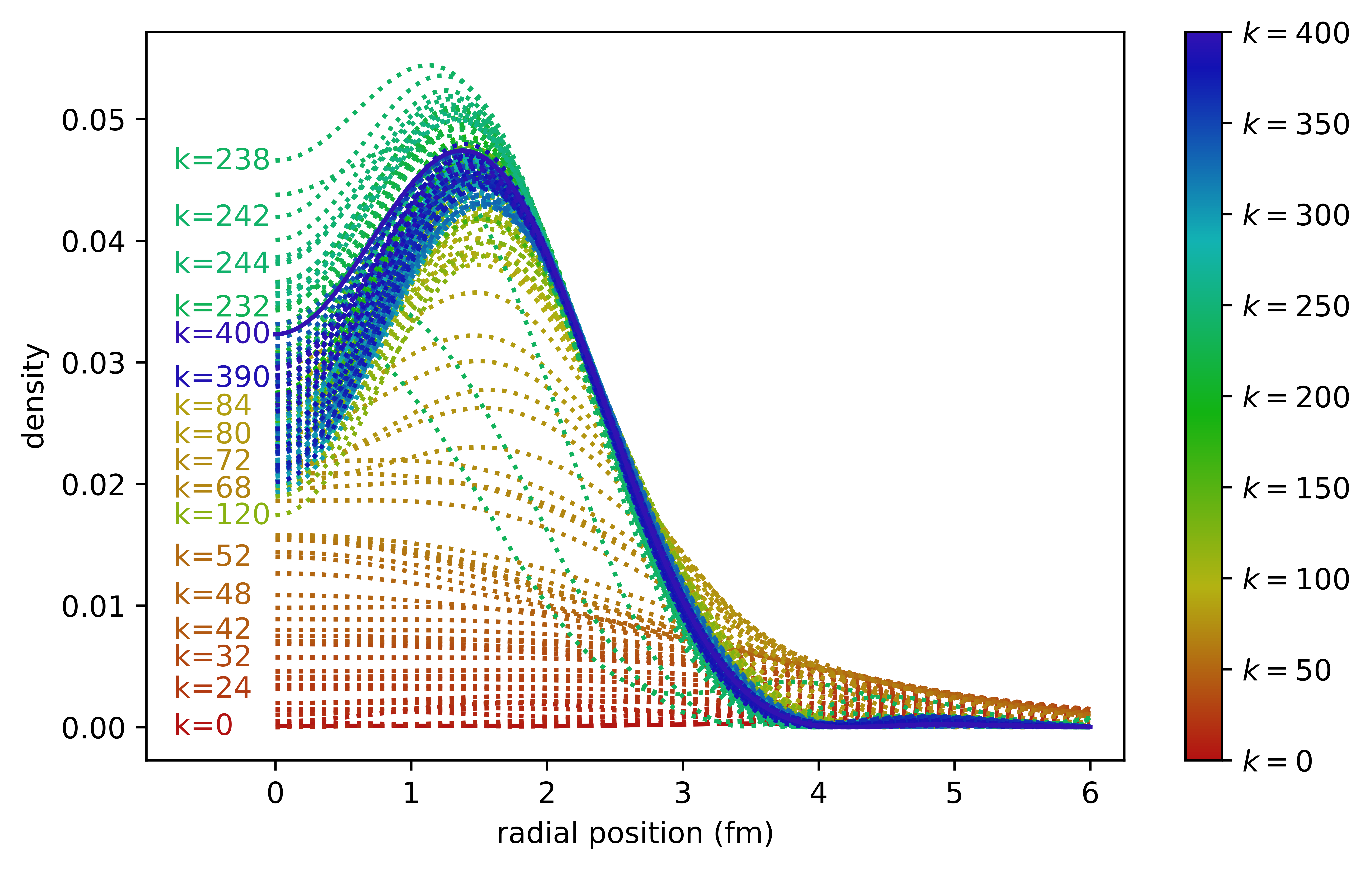}}
\end{figure}

\begin{figure}[H]
    \ContinuedFloat
    \subfloat[\label{OQm3_10k_diff}]{\includegraphics[width=\columnwidth]{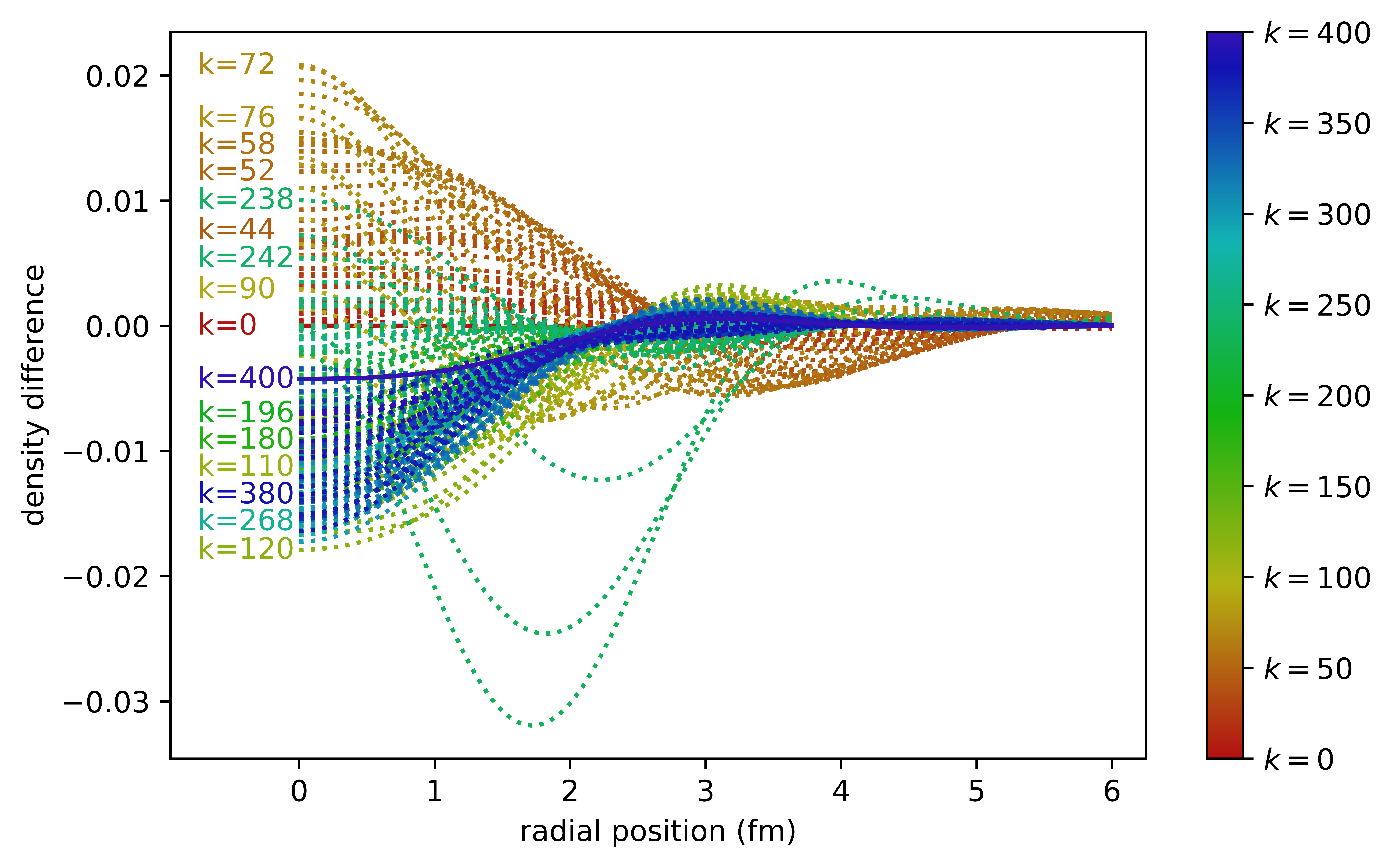}}
    
    \caption{ITE for $N=3$. (a) Classical ITE. (b) Quantum ITE with 10000 shots. (c) Difference between classical ITE and quantum ITE with 10000 shots.}
    \label{O3}
\end{figure}

Upon investigating, we find that some of the coefficients were registered to zero due to their small amplitudes, which significantly lowers their probabilities of being measured. As an initial attempt to improve this behaviour, the same quantum ITE is run with a higher sampling rate (100000 shots per measurement). Figure \ref{OQm3} shows the simulation result. Improvement of the unstable behaviour can be seen.

\begin{figure}[tbh]
    \subfloat[\label{OQm3_100k}]{\includegraphics[width=\columnwidth]{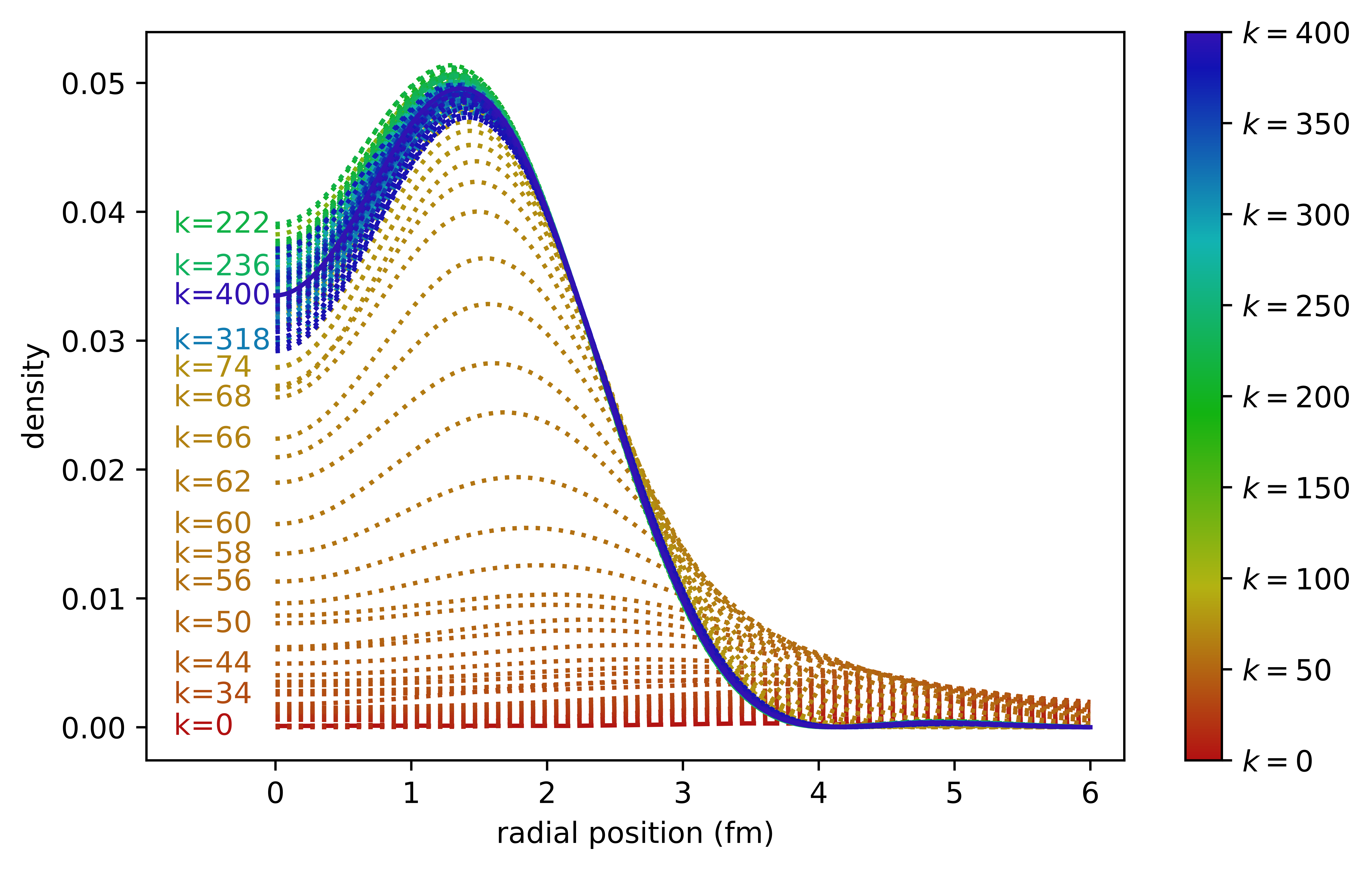}}
\end{figure}

\begin{figure}[H]
    \ContinuedFloat
    \subfloat[\label{OQm3_100k_diff}]{\includegraphics[width=\columnwidth]{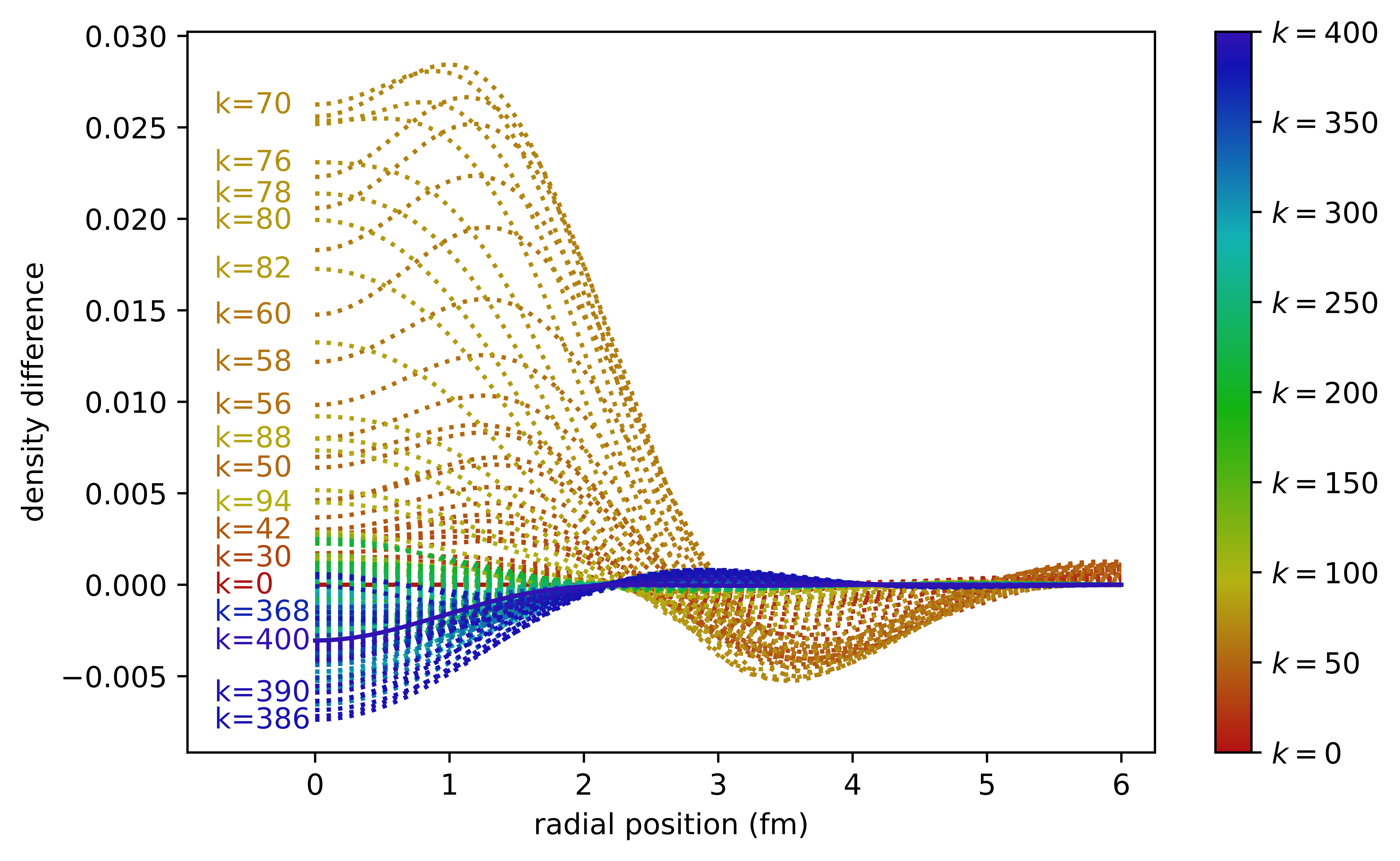}}
    
    \caption{ITE for $N=3$. (a) Quantum ITE with 100000 shots. (b) Difference between classical ITE and quantum ITE with 100000 shots.}
    \label{OQm3}
\end{figure}

The same behaviour appears in the $N=4$ case and affects more coefficients, as shown in figure \ref{O4}. With more coefficients failing to be measured in early time steps, the ITE of the state does not converge to the ground state. 

\begin{figure}[tbh]
    \subfloat[\label{OC4}]{\includegraphics[width=\columnwidth]{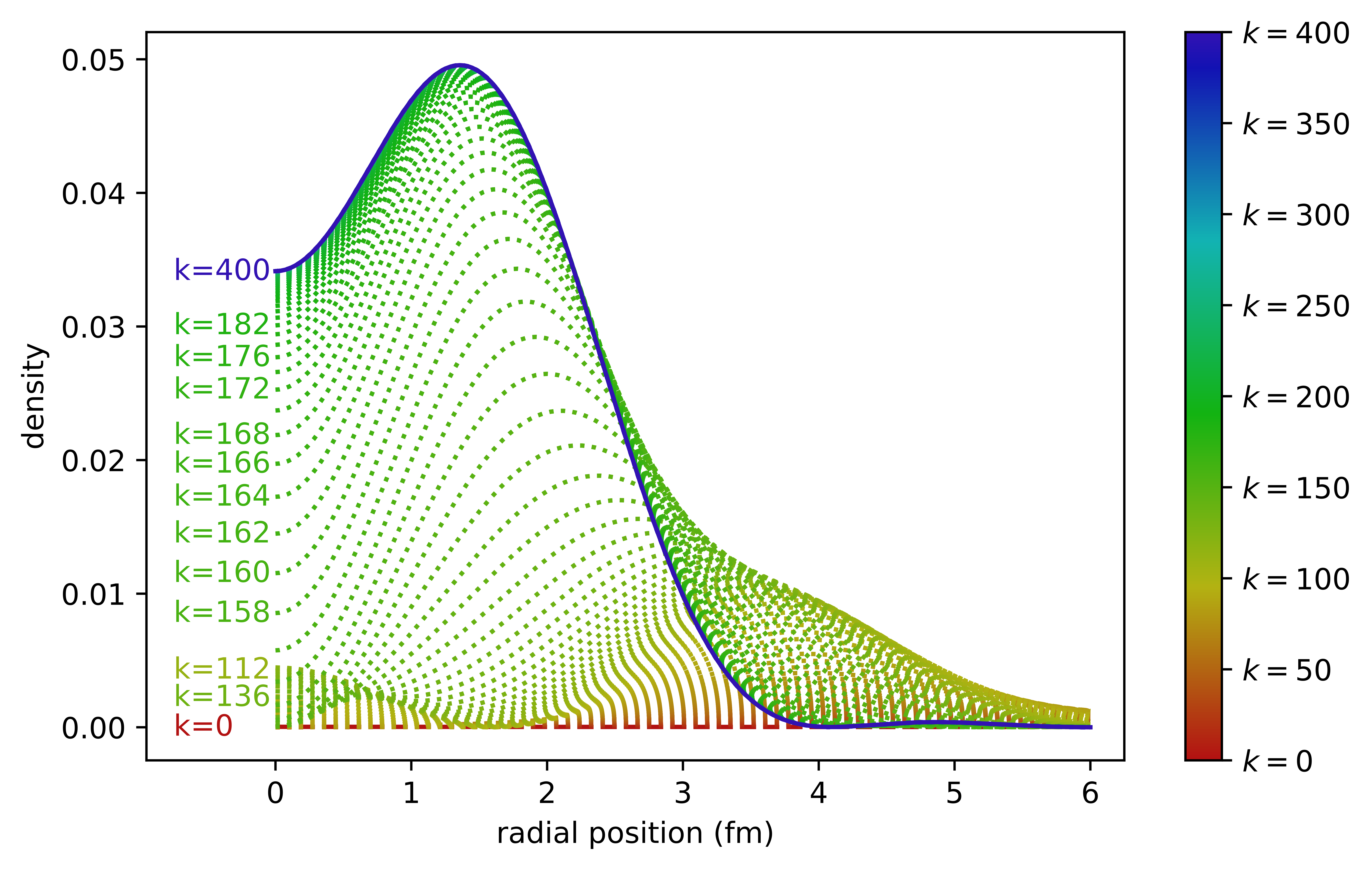}}
\end{figure}

\begin{figure}[H]
    \ContinuedFloat    
    \subfloat[\label{OQm4}]{\includegraphics[width=\columnwidth]{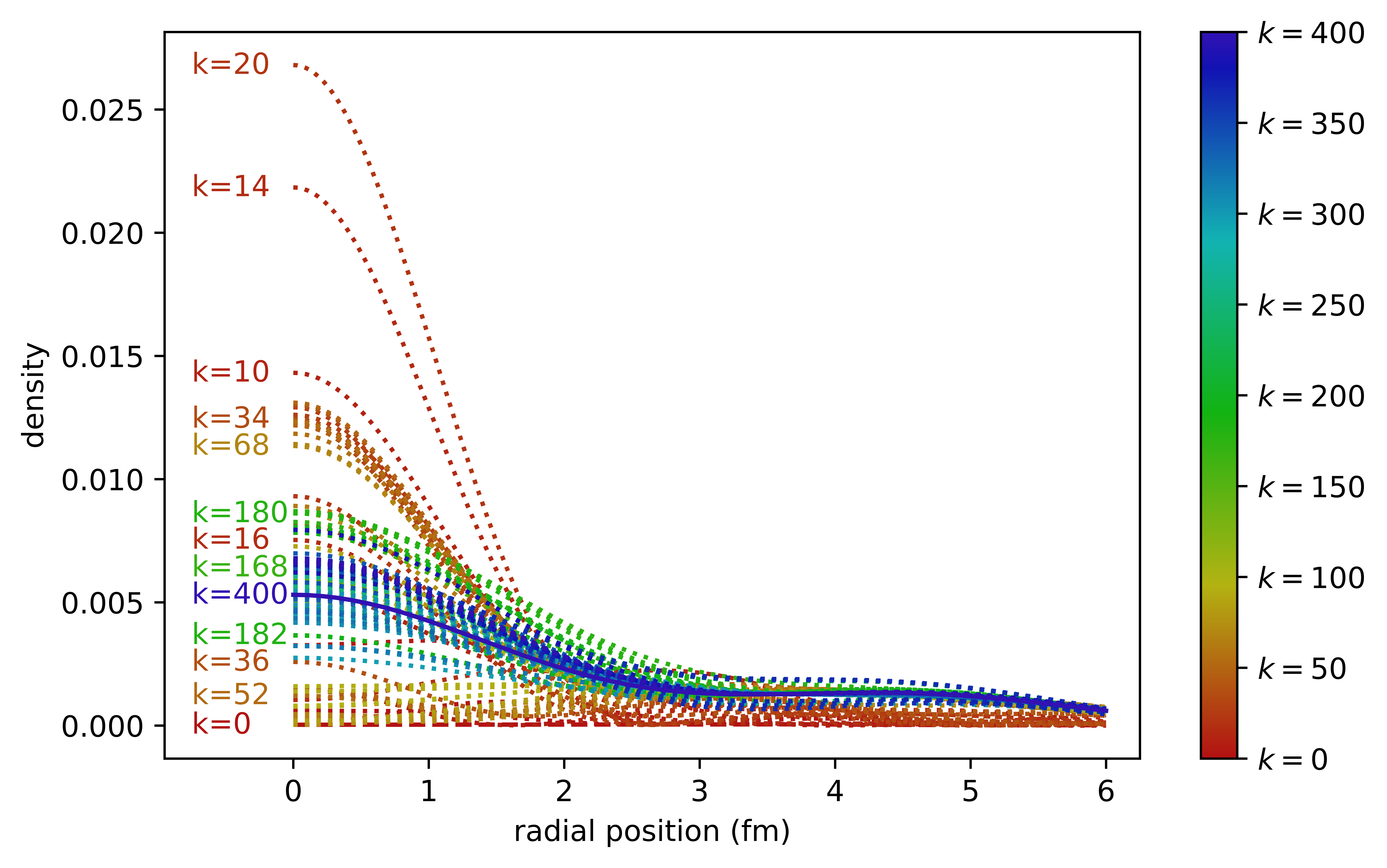}}
\end{figure}

\begin{figure}[H]
    \ContinuedFloat    
    \subfloat[\label{OQm4_diff}]{\includegraphics[width=\columnwidth]{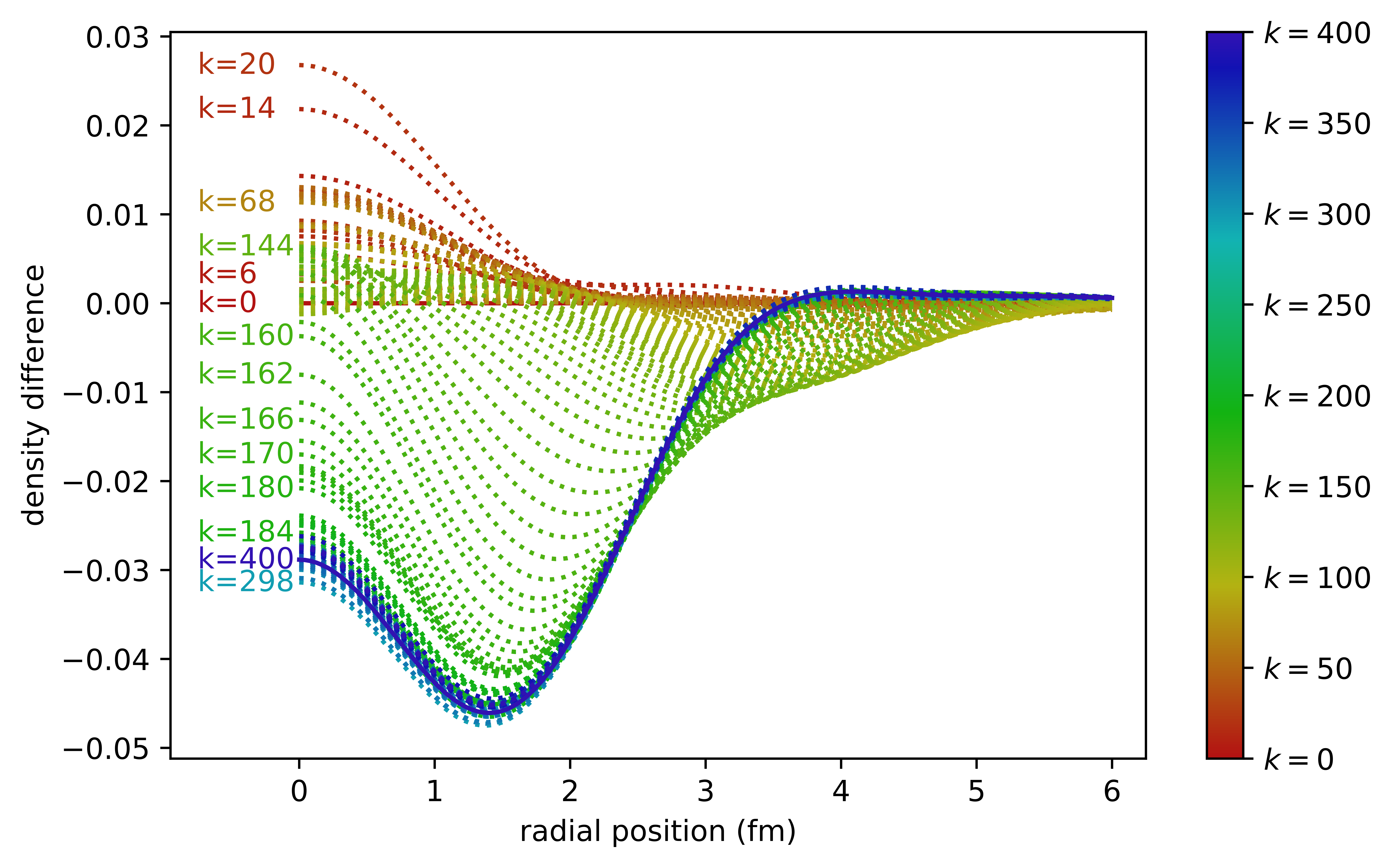}}
    
    \caption{ITE for $N=4$. (a) Classical ITE. (b) Quantum ITE with 10000 shots. (c) Difference between classical ITE and quantum ITE with 10000 shots.}
    \label{O4}
\end{figure}

It is not practically possible to increase the number of shots without limit, but we evaluate the "infinite shot" limit by running the simulation on the statevector backend from QISKIT \cite{MTreinish.2023}. This forms a check that the algorithm has been implemented faithfully.

\begin{figure}[tbh]
    \subfloat[\label{OQv4}]{
    \includegraphics[width=\columnwidth]{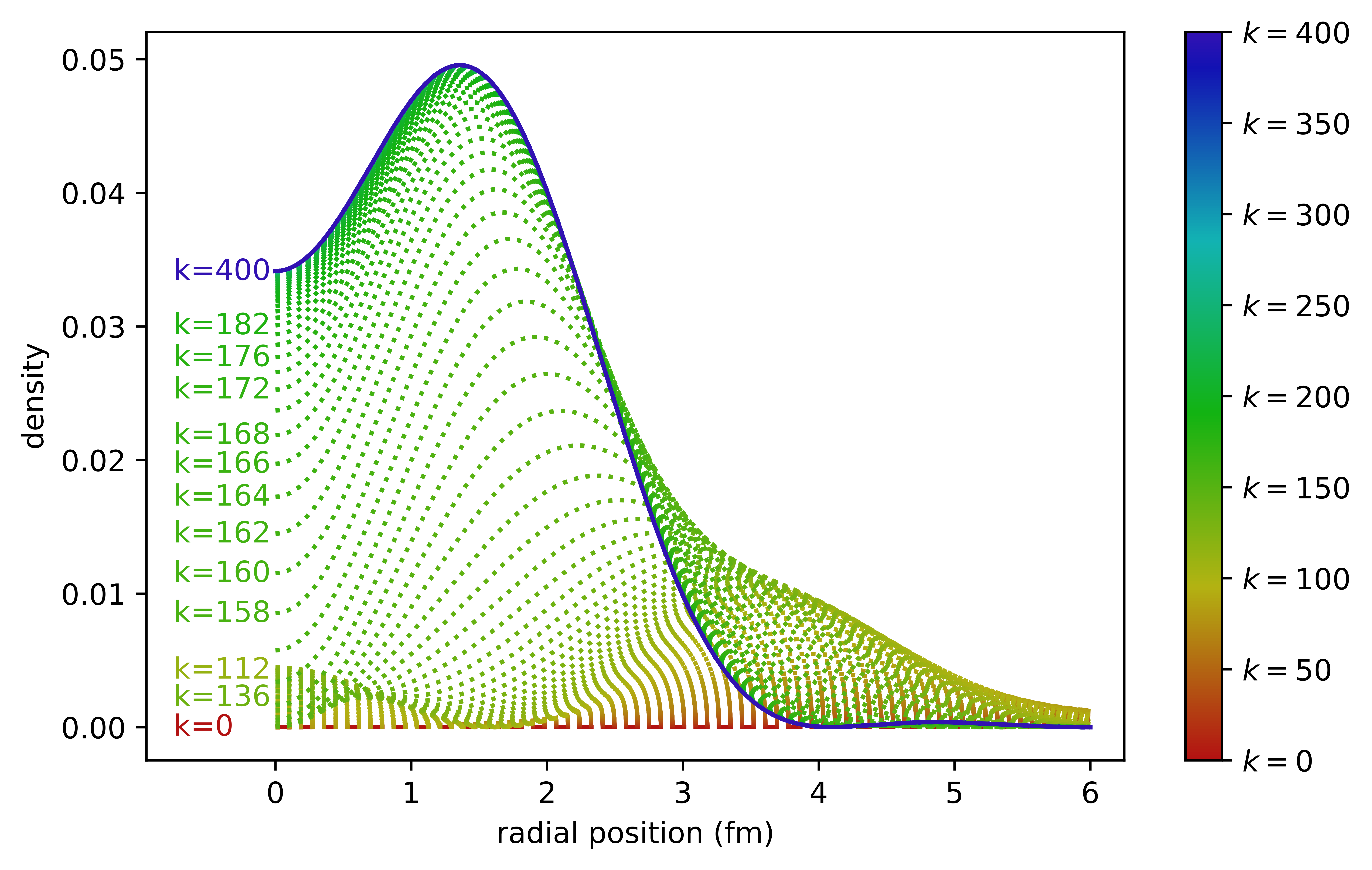}}
\end{figure}

\begin{figure}[H]
    \ContinuedFloat    
    \subfloat[\label{OQv4_diff}]{\includegraphics[width=\columnwidth]{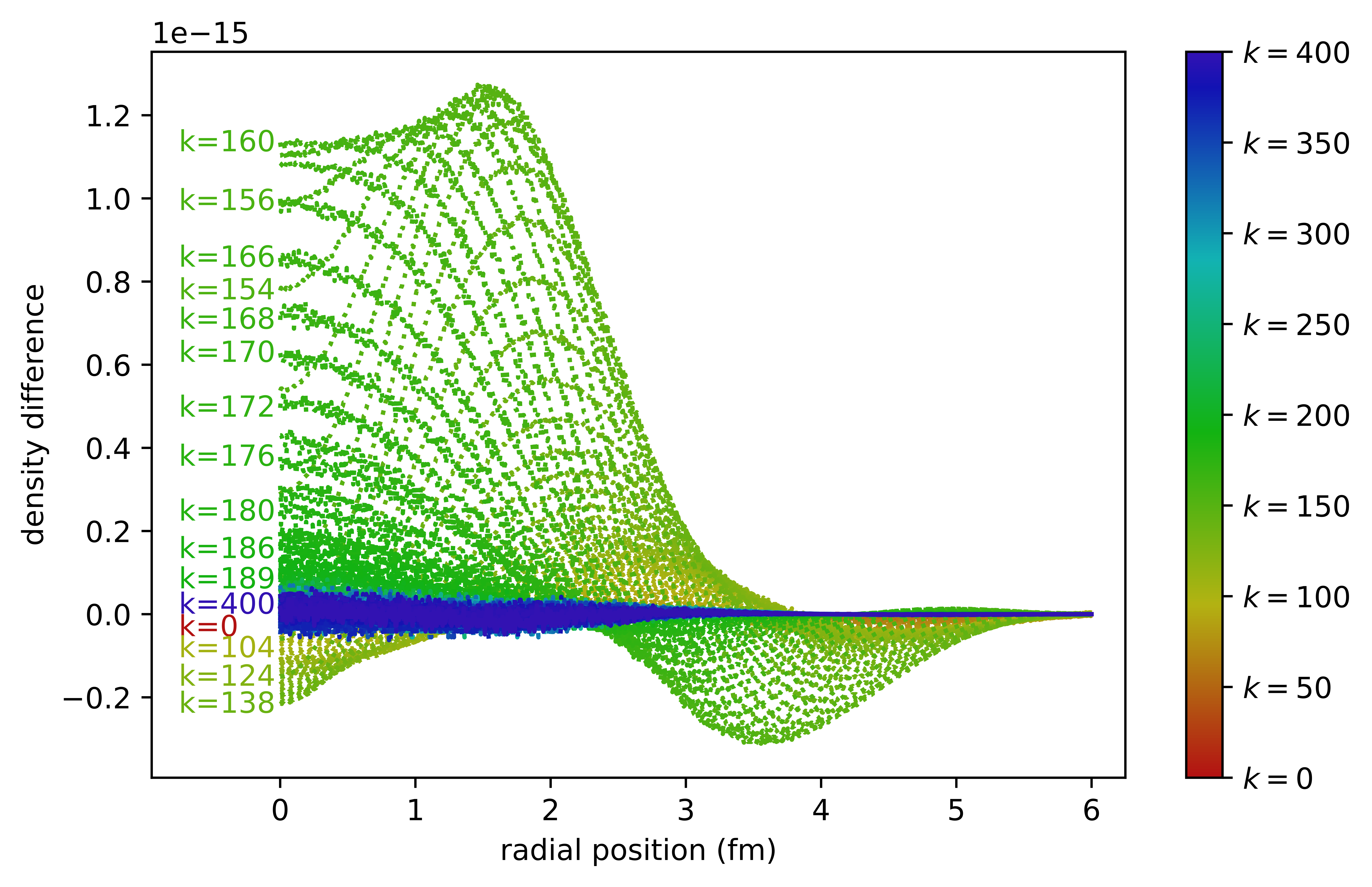}}    \caption{ITE for $N=4$. (a) Quantum ITE using statevector simulator. (b) Difference between classical ITE and quantum ITE using statevector simulator.}
    \label{OQ4}
\end{figure}

Figure \ref{OQ4} shows the results of this simulation. The per-iteration results and the ground state energy calculation are in close agreement with the classical algorithm.

\section{Limitations and Proposed Improvements}
Our algorithm makes use of an expanded Hilbert space by adding $2N$ ancillary qubits. A unitary operator in this expanded Hilbert space can then be chosen such that it corresponds to our desired non-unitary evolution operator in the target subspace \cite{CommunTheorPhys.45.825,Res.2020.1}. In theory only one ancillary qubit is needed for this implementation by adopting the quantum singular value transformation (QSVT) technique \cite{STOC2019.193}. The extra ancillary qubits in our implementation are traded for a reduced circuit depth. Previous QSVT implementations show scalings ranging from $O(2^N)$ \cite{Quantum.8.1311} to $O(4^N)$ \cite{arxiv:quant-ph/0504100}, while the corresponding subcircuit in our implementation scales as $O(N)$ \cite{PhysRevC.109.044322}. However, the complexity of the system still grows exponentially regardless of the implementation. This results in two exponentially growing computational bottlenecks in our algorithm.

\subsection{Success Probability}
In order to extract information in the subspace from the full Hilbert space, we require all the ancillary qubits to be measured to be zero. The probability for this (the success probability) determines the ratio of useful shots to the total number of shots. As the size of the circuit grows, the success probability decreases exponentially ($\sim O\left(2^{-2N}\right)$, with the exact form being Hamiltonian-dependent), and an exponentially increasing number of shots is required to maintain the same precision of the measurements of the expansion coefficients. The instability shown in figures \ref{OQm3_10k} and \ref{OQm4} originates from this.

An exponentially increasing shots requirement defeats the purpose of using a quantum algorithm for its scaling advantages. Techniques to mitigate low probability of shots leading to useful measurements have been enacted in other related methods of implementing the quantum imaginary time method \cite{Quantum.8.1311}. However, such a technique cannot immediately be applied in our case due to the imaginary-time-dependent potential caused by the non-linear Schr\"{o}dinger equation. In this section we propose an alternative solution to mitigate this limitation.

As seen in figure \ref{OQm4}, the state failed to evolve to the ground state due to errors in early time steps. Examination of the data reveals that the state is trapped in a local minimum, instead of evolving to the global minimum as expected. In response to this behaviour we use a larger imaginary time step. In classical ITE the imaginary time step is often optimized against and this is not a solution unique to our quantum algorithm. However, in our implementation the step size affects more than the converging rate. Increasing the step sizes results in more apparent changes in the coefficient values between iterations and prevent the state from being trapped.

\begin{figure}[tbh]
    \subfloat[\label{OCQ3_En_full}]{\includegraphics[width=\columnwidth]{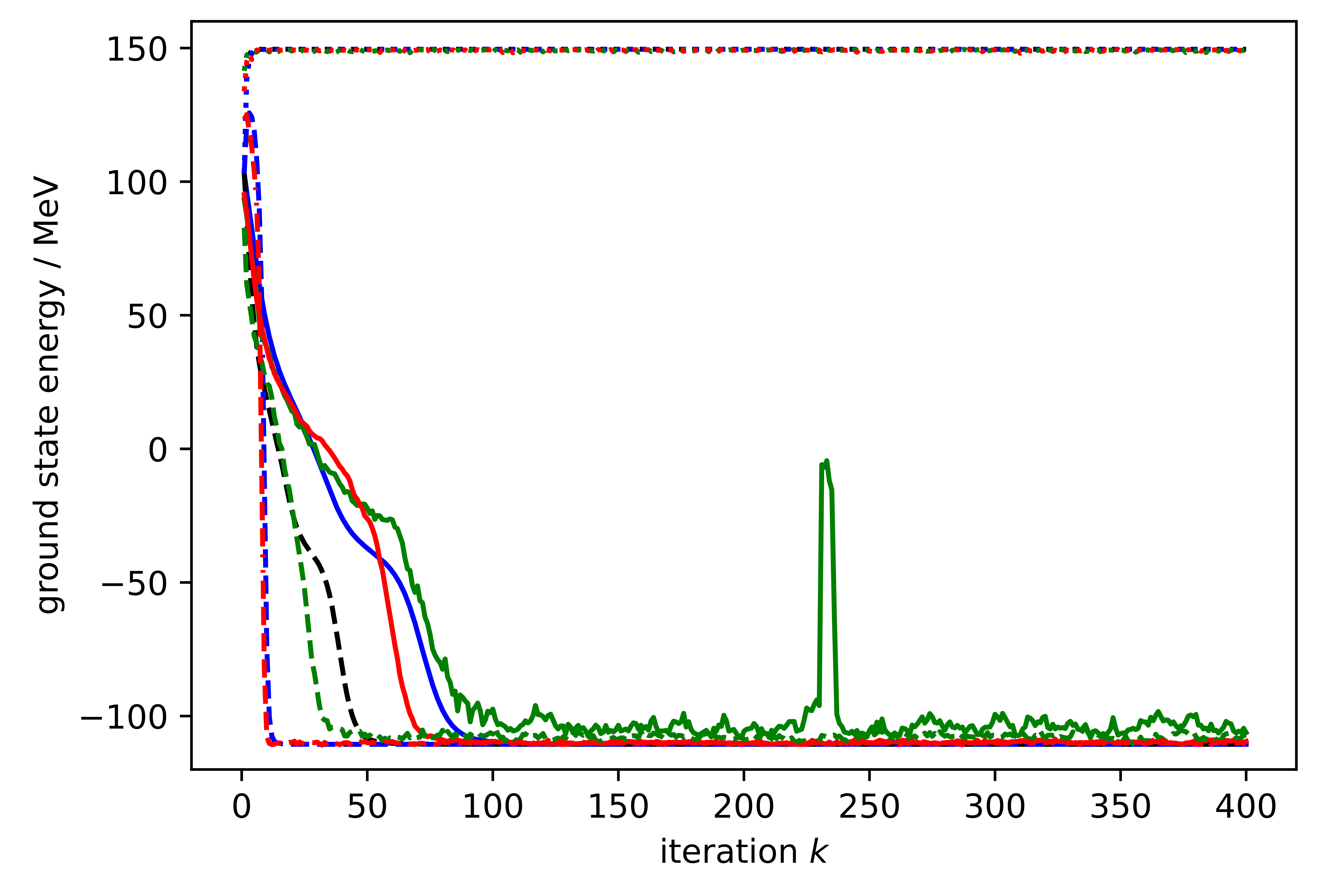}}
\end{figure}
\begin{figure}[H]
    \ContinuedFloat    
    \subfloat[\label{OCQ3_En_zoom}]{\includegraphics[width=\columnwidth]{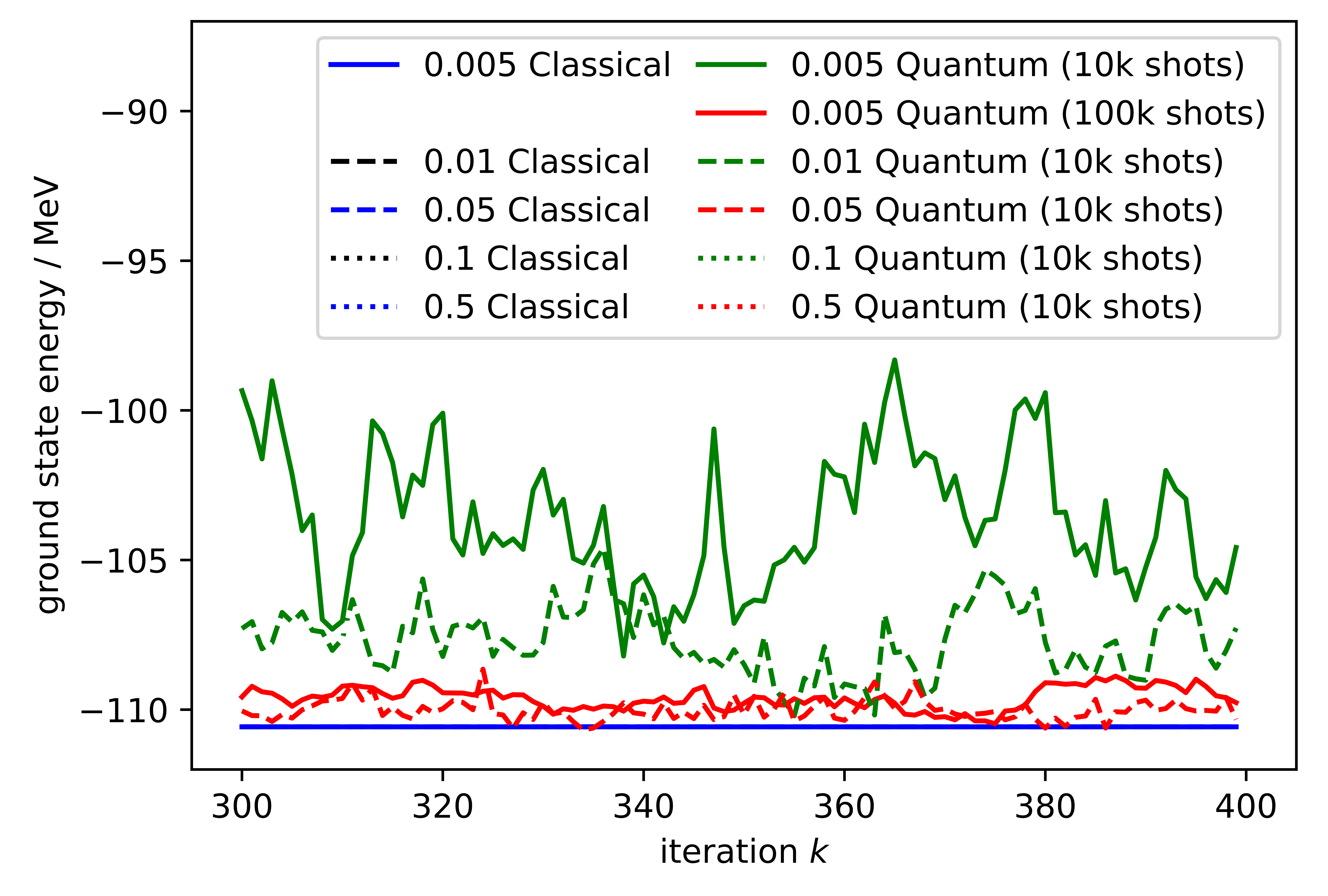}}
    \caption{Per iteration development of the ground state energy $E_{gs}$ for $N=3$ with varying step sizes. (a) Full view. (b) Zoomed in view.}
    \label{OCQ3_En}
\end{figure}

\begin{table}[tbh]
    \centering
    \begin{tabular}{cccc} 
        \hline\hline
        \multirow{2}{*}{$\frac{\Delta\tau}{\hbar}/\si{\MeV\tothe{-1}}$}&\multicolumn{3}{c}{$E_{gs}/\si{\MeV}$}\\
        &classical&10k shots&100k shots\\ \hline
        0.005&-110.57&-106.19&-109.47\\
        0.01&-110.57&-107.14&-110.30\\ 
        0.05&-110.57&-110.59&-110.62\\ 
        0.1&\textcolor{red}{149.55}&\textcolor{red}{149.24}&\textcolor{red}{149.47}\\
        0.5&\textcolor{red}{149.55}&\textcolor{red}{149.30}&\textcolor{red}{149.10}\\ \hline\hline
    \end{tabular}
    \caption{\ce{^{16}O} Hartree-Fock ground state energy after 400 ITE iterations with varying step sizes. Values obtained from failed convergence highlighted in red.}
    \label{step_tab}
\end{table}

Figure \ref{OCQ3_En} and table \ref{step_tab} show the ground state energies obtained after 400 iterations for the $N=3$ case with different step sizes. With an increasing imaginary time step, improvements of the unstable behaviour can be seen. However, too large an imaginary time step results in the approximations in equations (\ref{IntApprox}) and (\ref{eq:uapprox}) no longer holding. This is shown by failures of the ITE algorithm in both the classical and quantum cases for $\frac{\Delta\tau}{\hbar}=0.1\si{\MeV\tothe{-1}}$ and $\frac{\Delta\tau}{\hbar}=0.5\si{\MeV\tothe{-1}}$.

Thus, with an optimized imaginary time step, the success probability bottleneck can be mitigated without increasing the number of shots exponentially. Further improvement can be made by using an adaptive step size, but we have not explored the effects of that here.

\subsection{Circuit Size}
Another bottleneck lies in the state preparation sub-circuit $SP(2N)$, whose size scales as $O\left(2^{2N}\right)$ \cite{PhysRevC.109.044322, PhysRevA.83.032302} (see figure \ref{FlowChart} for scaling in different circuit parts). This originates from the preparation of the highly-entangled $2N$-qubit ancillary state $\ket{\phi_a}$, whose statevector contains the $2^{2N+1}-2$ independent coefficients. There has been multiple attempts for efficient state preparation algorithms \cite{IEEETransComput-AidedDesIntegrCircuitsSyst.43.161,PhysRevLett.129.230504,arxiv:2306.16831,QuantumSciTechnol.8.035027}. In this section we propose an alternative approach by using an approximate state of $\ket{\phi_a}$ that requires fewer quantum gates to set up.

In order to find such an approximate state $\ket{\phi'_a}$, we examine the evolution of the normalized decomposition coefficients $\beta^l_{i'}$ of the evolution of the non-unitary operator $\hat{O}^{l,(k)}$ (\ref{eq:uapprox}) in the Pauli basis \cite{PhysRevC.109.044322}, where $l$ denotes the corresponding $s/p$ states. The results are shown in figure \ref{PauliConrtibution}. The coefficients are grouped according to the weights of their corresponding Pauli strings $P_{i'}$. The identity gate $I$ is assigned a weight of 0 while each of the Pauli gate $X$, $Y$, and $Z$ has a weight of 1. This is equivalent to counting the number of $\sigma_P$ gates in $P_{i'}$. For example, the gate $IXIZ$ has a weight of 2 and the gate $ZYYX$ has a weight of 4.

\begin{figure}[tbh]
    \subfloat[\label{PauliMean}]{\includegraphics[width=\columnwidth]{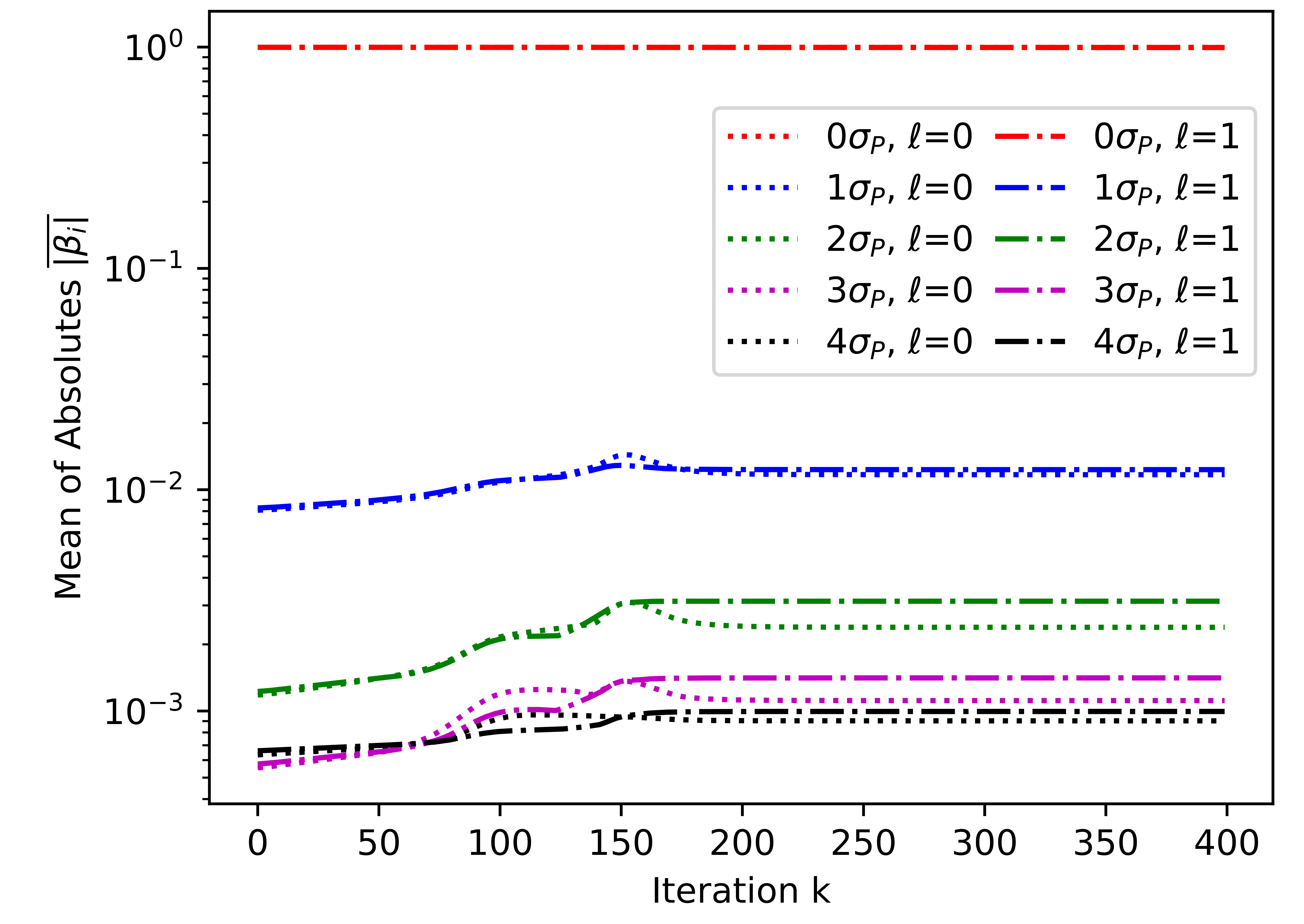}}
\end{figure}

\begin{figure}[tbh]
    \ContinuedFloat    
    \subfloat[\label{PauliSum}]{\includegraphics[width=\columnwidth]{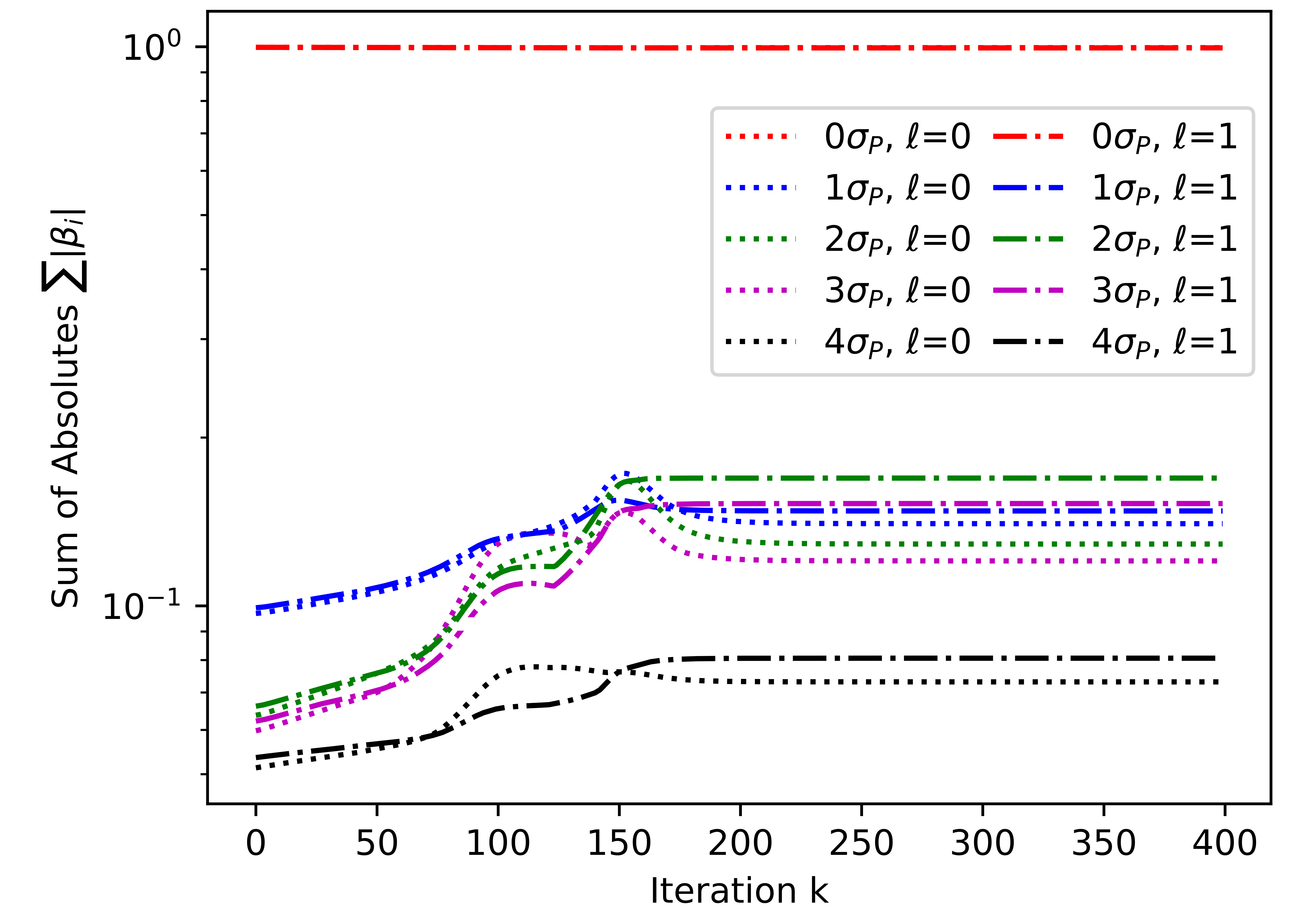}}
    \caption{Per-iteration development of the $\beta$ coefficients for the terms in the Pauli expansion of the unitary operator in the quantum imaginary time evolution algorithm for $N=4$ case, implemented with statevector simulator. Terms are grouped together based on the number of $X$, $Y$, or $Z$ gates included. (a) Mean of absolute values of $\beta$ coefficients. (b) Sum of absolute values of $\beta$ coefficients.}
    \label{PauliConrtibution}
\end{figure}

It can be seen that the contributions of operators decreases with weight, while the identity operator $IIII$ (weight 0) maintains an unperturbed contribution near 1.0 throughout the evolution. This relative importance of different terms allows for a basis truncation up to a certain threshold weight $w_t$. The number of independent coefficients then scales polynomially as $O\left(N^{w_t}\right)$ (see Appendix \ref{Trunc} for proof), which shows significant improvement from the exponential scaling $O\left(2^{2N}\right)$. Noticing the coefficients for any $P_{i'}$ containing an odd number of $Y$ gates is zero due to working only with real states, we can further reduce the number of independent variables by a factor of roughly half. (see Appendix \ref{Ypair})

Additionally, with the reduction of independent coefficients may allow for a more efficient use of ancillary qubits, which could further relieve the success probability bottleneck. However, this requires a restructuring of the algorithm and may lead to increased circuit depth in the operator sub-circuit, which is out of the scope of the work presented here.

\section{Conclusion}
We have presented an algorithm solving coupled non-linear Schr\"{o}dinger equations via the imaginary time evolution method. Using Hartree-Fock equations for the oxygen-16 nucleus as an example, we show that this implementation provides results in agreement with the classical algorithm. With a larger basis, however, our current algorithm faces its limitations.

On the one hand, some of the coefficients failed to be detected with our number of measurements. We show, with a statevector simulation, that the limitation lies not within the circuit, but the measurements. As a preliminary measure, we improve this by increasing the number of shots in the $N=3$ case. We further show that an increase in the imaginary time step helps mitigate the problem without increasing computational cost. This paves the way for the use of an optimized or even adaptive step size for a future implementation.

On the other hand, the ancillary state preparation sub-circuit is proved to be the bottleneck of our algorithm with its exponentially growing circuit size. By investigating the relative contributions of different Pauli strings $P_{i'}$ from the decomposition of the ITE operator $\hat{O}^{(k)}$ in the Pauli basis, we propose a basis truncation which could reduce the scaling to a polynomial one.

\section*{Declarations}

The authors declare no competing interests.

\section*{Acknowledgements}
We acknowledge funding from the UK Science and Technology Facilities Council (STFC) under grant numbers ST/V001108/1 and ST/W006472/1, and from SEPNET.

\section*{Data Availability Statement}

Data sets generated during the current study are available from the corresponding author on reasonable request.

\begin{appendices}

\section{Non-linearity and Iteration}\label{NlH}
The Hamiltonian we use (equation \ref{Ham}) is non-linear and dependent on the states we are solving for (through the particle density),
\begin{equation}
    \hat{H}^l_{\text{hf}}=\hat{T}^l+V\left(\psi_l\right)\text{.}
\end{equation}

\begin{figure*}[p!]
    \centering
    \includegraphics[width=\textwidth]{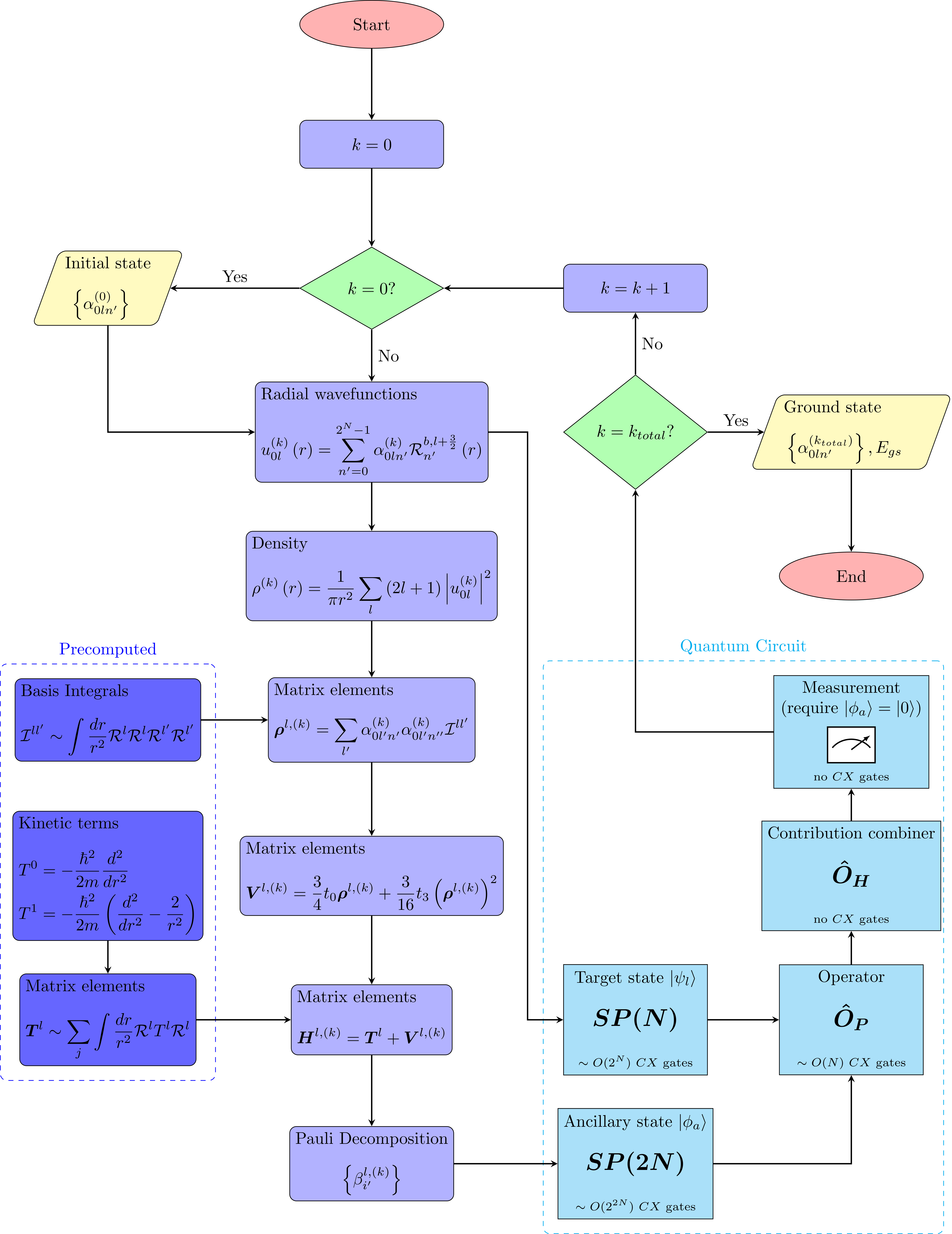}
    \caption{A flow chart showing the procedure of the hybrid QITE algorithm. Classical parts are denoted by blue boxes with rounded-corners and quantum parts are denoted by cyan boxes with pointy corners.}
    \label{FlowChart}
\end{figure*}

The ITE of a state $\ket{\psi\left(\tau\right)}$ for an infinitesimal imaginary time step $d\tau$ from $\tau=\tau_0$ to $\tau=\tau_0+d\tau$ is given by 
\begin{equation}\label{imev}
    \ket{\psi\left(\tau_0+d\tau\right)}=\mathcal{N}\exp[-\frac{\hat{H}\left(\psi\left(\tau_0\right)\right)}{\hbar}d\tau]\ket{\psi\left(\tau_0\right)}\text{,}
\end{equation}
where the Hamiltonian shows an imaginary time dependence implicitly through the state $\ket{\psi\left(\tau\right)}$.

The ITE process is then discretised into $k_{total}$ steps, each with an imaginary time step of $\Delta\tau$. Equation \ref{imev} can be written as
\begin{equation}\label{itev}
    \ket{\psi^{(k+1)}}=\mathcal{N}\exp[-\frac{\hat{H}^{(k)}}{\hbar}\Delta\tau]\ket{\psi^{(k)}}\text{,}
\end{equation}
where
\begin{align}
    \hat{H}^{(k)}&=\hat{H}\left(\psi\left(k\Delta\tau\right)\right),\\
    \ket{\psi^{(k)}}&=\ket{\psi\left(k\Delta\tau\right)}\text{.}
\end{align}
At $k=0$, $\ket{\psi^{(0)}}=\ket{\psi\left(0\right)}$ is the initial state. The procedure of this iterative method is summarised in figure \ref{FlowChart}.

\section{State Preparation Subcircuits}\label{SP}
Denoting the $2^\mathfrak{N}$ coefficients in a $\mathfrak{N}$-qubit real state as $\{c_i\}$, $i=0,1,\dots,2^\mathfrak{N}-1$, the $\mathfrak{N}$-qubit state preparation subcircuit $SP(\mathfrak{N})$ is given in figure \ref{SPN},\\
\begin{center}
    \includegraphics[width=\columnwidth]{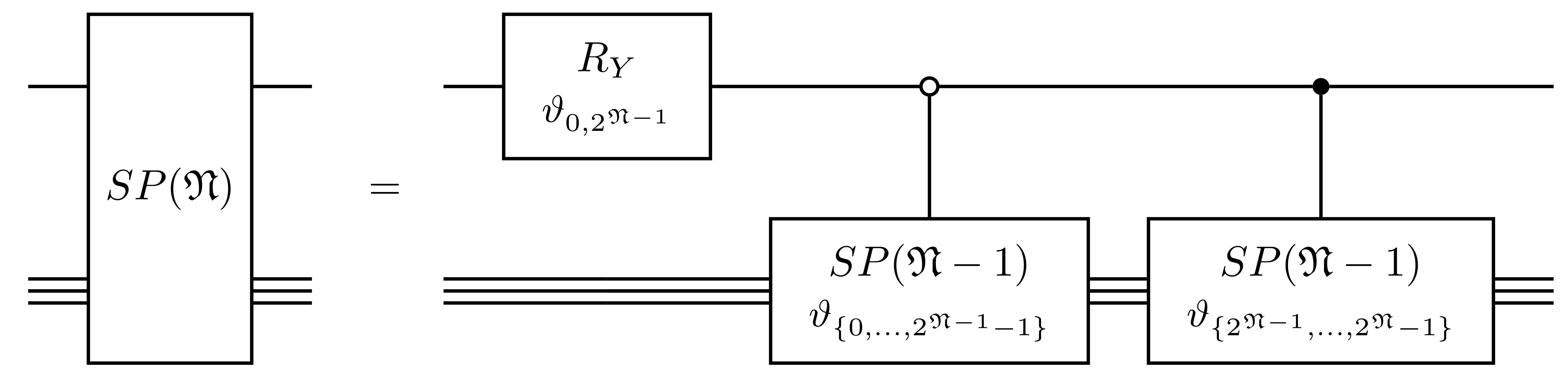}
    \captionof{figure}{$\mathfrak{N}$-qubit real state preparation.}
    \label{SPN}
\end{center}
where the angles of rotation, $\vartheta_{ij}$, for the state preparation, are given by
\begin{equation}
    \tan{\frac{\vartheta_{ij}}{2}}=\sqrt{\frac{\sum_{i'=j}^{2j-i-1}c_{i'}^2}{\sum_{i'=i}^{j-1}c_{i'}^2}}\text{,}
\end{equation}
and $SP(1)$ is given in figure \ref{SP1}.\\
\begin{center}
    \includegraphics[width=\columnwidth]{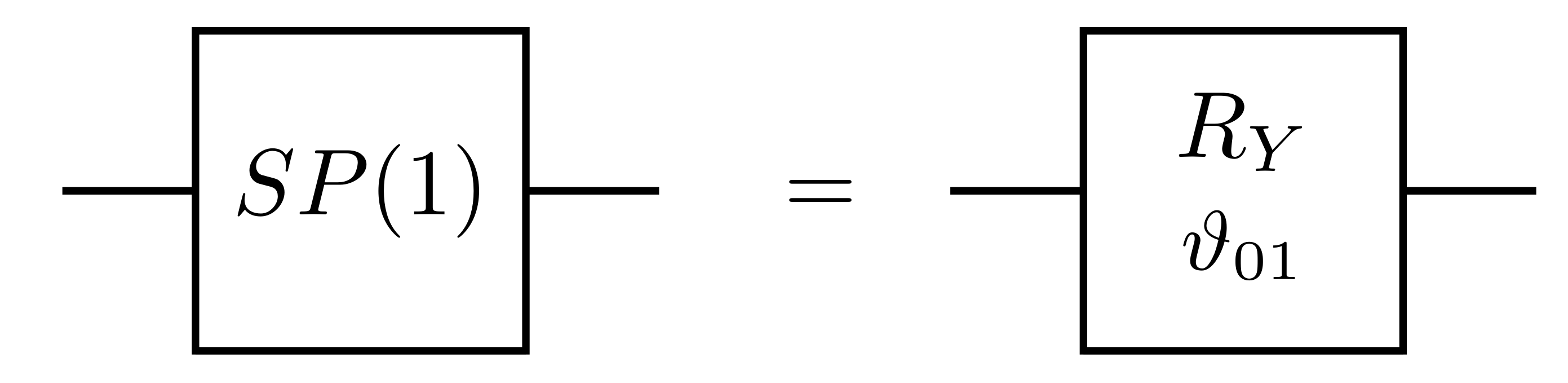}
    \captionof{figure}{1-qubit real state preparation.}
    \label{SP1}
\end{center}

\section{Scaling of Truncated Basis}\label{Trunc}
With a basis truncation at threshold weight $w_t$, the total number of terms in the truncated decomposition is given by
\begin{equation}\label{F_form}
    F\left(N,w_t\right)=\sum_{w=0}^{w_t}\binom{N}{w}3^w\text{.}
\end{equation}
This sum does not have a closed form but we can put a bound onto its scaling.

For any binomial coefficient $\binom{N}{w}$ we have
\begin{equation}
    \binom{N}{w}=\frac{N(N-1)\cdots[N-(w-1)]}{w!}\leq\frac{N^w}{w!}
\end{equation}

Thus,
\begin{equation}
    F\left(N,w_t\right)=\sum_{w=0}^{w_t}3^w\binom{N}{w}\leq\sum_{w=0}^{w_t}\frac{\left(3N\right)^w}{w!}\text{.}
\end{equation}

Hence, we have shown that the scaling of $F\left(N,w_t\right)$ is bounded by a polynomial scaling
\begin{equation}
    F\left(N,w_t\right)\leq\sum_{w=0}^{w_t}\frac{\left(3N\right)^w}{w!}\sim O\left(N^{w_t}\right)\text{.}
\end{equation}

\section{Scaling of Reduced $Y$ terms}\label{Ypair}
Ignoring terms with an odd number of $Y$ gates, on top of a basis truncation at threshold weight $w_t$, the remaining number of terms is
\begin{equation}
    G\left(N,w_t\right)=\sum_{w=0}^{w_t}\binom{N}{w}g_e\left(w\right)\text{,}
\end{equation}
where $g_e\left(w\right)$ is given by
\begin{equation}
    g_e\left(w\right)=\sum_{w'=0}^{\left\lfloor\frac{w}{2}\right\rfloor}\binom{w}{2w'}2^{w-2w'}\text{.}
\end{equation}

Constructing
\begin{equation}
    g_o\left(w\right)=\sum_{w'=0}^{\left\lfloor\frac{w-1}{2}\right\rfloor}\binom{w}{2w'+1}2^{w-\left(2w'+1\right)}\text{,}
\end{equation}

\noindent we have
\begin{equation}\begin{aligned}
    g_e\left(w\right)+g_o\left(w\right)    &=\sum_{w''=0}^{w}\binom{w}{w''}2^{w-w''}\left(1\right)^{w''}\\
    &=\left(2+1\right)^w=3^w\text{,}
\end{aligned}\end{equation}
and
\begin{equation}\begin{aligned}
    g_e\left(w\right)-g_o\left(w\right)    &=\sum_{w''=0}^{w}\binom{w}{w''}2^{w-w''}\left(-1\right)^{w''}\\
    &=\left(2-1\right)^w=1\text{.}
\end{aligned}\end{equation}

Hence, $g_e\left(w\right)$ and $G\left(N,w_t\right)$ are given by
\begin{equation}
    g_e\left(w\right)=\frac{3^w+1}{2}
\end{equation}
and
\begin{equation}\label{G_form}
    G\left(N,w_t\right)=\sum_{w=0}^{w_t}\binom{N}{w}\frac{3^w+1}{2}\text{.}
\end{equation}

Comparing the form of $G\left(N,w_t\right)$ (\ref{G_form}) against $F\left(N,w_t\right)$ (\ref{F_form}), we can see that omitting the odd $Y$ terms reduces the number of terms by about half.

\end{appendices}


\bibliography{sn-bibliography}

\end{document}